\documentclass[aps, prb, twocolumn]{revtex4}
\usepackage{graphicx}
\usepackage{enumerate}
\usepackage{hyperref}
\usepackage{textcomp}
\usepackage{amsmath}
\usepackage{amssymb}
\usepackage{pgf}
\usepackage{diagbox}
\usepackage{booktabs}
\usepackage{bbold}
\usepackage{color}
\usepackage{bbm}
\usepackage{soul}

\def\bra#1{\langle#1 |}
\def\ket#1{| #1\rangle}
\newcommand{\braket}[2]{\langle #1 | #2 \rangle}

\def\ud{\mathrm{d}}
\def\ee{\mathrm{e}}

\def\op#1{\hat{#1}}

\newcommand{\QA}{\mathrm{\scriptscriptstyle QA}}

\newcommand{\bomega}{\boldsymbol{\omega}}
\newcommand{\regular}{\mathrm{reg}}
\newcommand{\tcomp}{t_\mathrm{cc}}

\DeclareSymbolFont{matha}{OML}{txmi}{m}{it}
\DeclareMathSymbol{\varv}{\mathord}{matha}{118}

\newcommand{\bvec}{\boldsymbol{v}}

\newcommand{\bbeta}{\boldsymbol{\beta}}
\newcommand{\bgamma}{\boldsymbol{\gamma}}

\newcommand{\bpauli}{\hat{\boldsymbol{\sigma}}}

\newcommand{\calO}{{\mathcal{O}}}

\newcommand{\calK}{{\mathcal{K}}}
\newcommand{\calS}{{\mathcal{S}}}
\newcommand{\Top}{\widehat{{\mathcal{T}}}_{\PBC}}
\newcommand{\Ttwistop}{\widehat{{\mathcal{T}}}_{\ABC}}
\newcommand{\calP}{{\mathcal{P}}}

\newcommand{\gs}{\mathrm{gs}}

\newcommand{\pauli}{\op{\sigma}}
\newcommand{\PauliSigma}{\hat{\sigma}}

\newcommand{\bpaulitau}{{\hat{\boldsymbol{\tau}}}}

\newcommand{\versorz}{{\hat{\boldsymbol{z}}}}

\newcommand{\versorb}{{\hat{\boldsymbol{b}}}}

\newcommand{\versorom}{\hat{\boldsymbol{\omega}}}

\newcommand{\PBC}{{\scriptscriptstyle \mathrm{PBC}}}
\newcommand{\ABC}{{\scriptscriptstyle \mathrm{ABC}}}
\newcommand{\smallpm}{{\scriptscriptstyle (\pm)}}
\newcommand{\smallp}{{\scriptscriptstyle (+)}}
\newcommand{\smallm}{{\scriptscriptstyle (-)}}

\newcommand{\Texp}{\mathcal{T} \mathrm{exp}}
\newcommand{\Tprod}[1]{\prod^{\leftarrow #1}}

\newcommand{\res}{\mathrm{res}}

\newcommand{\opt}{\mathrm{opt}}

\newcommand{\sgdigital}{\scriptscriptstyle\mathrm{step}}
\newcommand{\mgdigital}{\scriptscriptstyle\mathrm{digit}}
\newcommand{\effective}{\mathrm{eff}}

\newcommand{\ceil}[1]{\left\lceil #1 \right\rceil}

\newcommand{\Ham}{\widehat{H}}
\newcommand{\Uevol}{\widehat{U}}
\newcommand{\Rrot}{\mathcal{R}}

\newcommand{\eres}{\epsilon^{\res}}
\newcommand{\Tann}{\tau}
\newcommand{\Ptrot}{\mathrm{P}}
\newcommand{\kconn}{\mathrm{k}}
\newcommand{\Hred}{\hat{\mathcal H}}
\newcommand{\Parity}{\calP}
\newcommand{\Parityop}{\widehat{\Parity}}
\newcommand{\Nred}{N_{\scriptscriptstyle \mathrm{R}}}

\begin{document}

\title{Quantum Annealing: a journey through Digitalization, Control, and hybrid Quantum Variational schemes}

\author{Glen Bigan Mbeng}
\affiliation{SISSA, Via Bonomea 265, I-34136 Trieste, Italy}
\affiliation{INFN, Sezione di Trieste, I-34136 Trieste, Italy}
\author{Rosario Fazio}
\affiliation{Abdus Salam ICTP, Strada Costiera 11, 34151 Trieste, Italy}
\affiliation{Dipartimento di Fisica, Universit\`a di Napoli ``Federico II'', Monte S. Angelo, I-80126 Napoli, Italy}
%
\author{Giuseppe E. Santoro}
\affiliation{SISSA, Via Bonomea 265, I-34136 Trieste, Italy}
\affiliation{Abdus Salam ICTP, Strada Costiera 11, 34151 Trieste, Italy}
\affiliation{CNR-IOM Democritos National Simulation Center, Via Bonomea 265, I-34136 Trieste, Italy}

\begin{abstract}
We establish and discuss a number of connections between a digitized version of Quantum Annealing (QA) with the Quantum Approximate Optimization Algorithm (QAOA) 
introduced by Farhi {\em et al.} (arXiv:1411.4028) as an alternative hybrid quantum-classical variational scheme for quantum-state preparation and optimization.
We introduce a technique which allows to prove, for instance, a rigorous bound concerning the performance of QAOA for MaxCut on a $2$-regular graph, 
equivalent to an unfrustrated antiferromagnetic Ising chain. 
The bound shows that the optimal variational error of a depth-$\Ptrot$ quantum circuit has to satisfy $\eres_{\Ptrot}\ge (2\Ptrot+2)^{-1}$. 
In a separate work (Mbeng {\em et al.}, arXiv:1911.12259) we have explicitly shown, exploiting a Jordan-Wigner transformation, that among the $2^{\Ptrot}$ degenerate 
variational minima which can be found for this problem, all strictly satisfying the equality $\eres_{\Ptrot}=(2\Ptrot+2)^{-1}$, one can construct a special 
{\em regular} optimal solution,
which is computationally optimal and does not require any prior knowledge about the spectral gap. 
We explicitly demonstrate here that such a schedule is adiabatic, in a digitized sense, and can therefore be interpreted as an optimized digitized-QA protocol.  
We also discuss and compare our bound on the residual energy to well-known results on the Kibble-Zurek mechanism behind a continuous-time QA.    
These findings help elucidating the intimate relation between digitized-QA, QAOA, and optimal Quantum Control. 
%
\end{abstract}

\maketitle

\section{Introduction} \label{sec:intro}
%
In the past two decades there have been great advances in the field of quantum computation~\cite{Nielsen_Chuang:book}.
However, in spite of the steady experimental progress of quantum technologies, the realization of a quantum computer capable of reliably running quantum algorithms 
which provably outperform their classical counterparts is still out of reach. 
State of art quantum devices are only prototypical quantum computers and broadly belong to the class of Noisy Intermediate-Scale Quantum (NISQ) technologies \cite{Preskill_Quantum2018}. 
Developing and improving algorithms which are suitable to run on NISQ technologies is essential to achieve quantum speedup~\cite{Ronnow2014} or quantum supremacy~\cite{Martinis_Nat2019}. 
Promising candidates in this field are {\em quantum annealing} and {\em hybrid variational schemes}, which aim at solving specific optimization problems better than available classical algorithms.

Quantum Annealing (QA)~\cite{Finnila_CPL94, Kadowaki_PRE98, Brooke_SCI99, Santoro_SCI02},  
{\em alias} Adiabatic Quantum Computation~\cite{Farhi_SCI01, Albash_RevModPhys2018},  
is as an effective way for finding solutions to optimization problems by running a continuous time evolution on special purpose analog quantum devices, named quantum annealers. 
The run (or annealing) time of the algorithm, $\tau$, is associated with the smallest spectral gap encountered during the annealing dynamics. 
In particular, vanishing spectral gaps constitute a bottleneck for the performance of QA. 
This is especially severe in cases where a first-order transition, accompanied by an exponentially small spectral gap, must be crossed to go from the initial state
to the target solution. 
Various strategies have been proposed to address this issue, including heuristic guesses for the initial state~\cite{Perdomo_QIP2011}, 
strategies for increasing the minimum gap~\cite{Zeng_JPA2016, Zhuang_PRA14} or avoiding first-order lines~\cite{Seoane_JPA2012}, and the quantum 
adiabatic brachistochrone formulation~\cite{Zanardi_PRL2009}. 
In some cases, modifying the annealing protocol is a must for a quantum speedup~\cite{Roland_PRA2002}. 
However, this is usually~\cite{Zeng_JPA2016, Zhuang_PRA14,Seoane_JPA2012,Zanardi_PRL2009,Roland_PRA2002} done by exploiting an 
{\em a priori} knowledge on the spectral gap, which is in general unknown or hard to access~\cite{Ambainis_arXiv2013, Wolf_Nat2015}.

Hybrid variational schemes use parameterized digital quantum circuits to approximate the solution of hard optimization problems~\cite{Farhi_arXiv2014,Peruzzo_NatComm14,Zoller_NAT19}.  
In these algorithms, a classical minimization routine optimizes the variational parameters of the quantum circuit and returns the best approximate solution. 
Although each hybrid variational scheme is tailored to solve a specific class of problems, they can all conveniently run on general purpose digital quantum devices. 
The Quantum Approximate Optimization Algorithm (QAOA)~\cite{Farhi_arXiv2014} is a popular hybrid variational scheme designed specifically to solve combinatorial 
optimization problems. It operates through a depth-$\Ptrot$ circuit of digitized unitary gates. 
In the QAOA framework, the run time of the algorithm is related to the depth $\Ptrot$ of the variational quantum circuit and to the number of iterations that the classical 
minimization routine requires to find good variational parameters. 
 
QA and QAOA are two universal models of computation~\cite{Ahronov_2004,Lidar_PRL07,Lloyd_arXiv2018}, designed to run on different NISQ hardware and subject to 
different limitations. However, Ref.~\onlinecite{Martinis_Nat16} and Ref.~\onlinecite{Yang_PRX2017} suggest a possible connection between these two models. 
On one hand, Ref.~\onlinecite{Martinis_Nat16} has advocated using digitized-QA (dQA), a fully digital version of QA that runs on general purpose digital quantum devices 
and allows the possibility of doing error-correction~\cite{Kitaev_arXiv1998, Austin_PRA2012}. 
On the other hand, in the field of optimal quantum control, Yang {\em et al.}~\cite{Yang_PRX2017} have used Pontryagin's 
principle~\cite{Dalessandro2007, Brif_NewJPhys2010} to shown that QAOA circuits are sufficient to implement time optimal protocols for ground state preparation.

Here, and in the related paper Ref.~\onlinecite{Mbeng_optimal-dQA_arXiv2019}, we close the loop, by making a step forward in establishing this connection. 
We study the performance of dQA and QAOA in a unified framework, and we show that one can construct optimal QAOA solutions which are inherently 
{\em adiabatic}, in a digitized-QA framework \cite{Martinis_Nat16}. 

More in detail, our paper focuses on two main results.
The first, methodological, is a technique to establish a variational bound on the residual energy of a class of antiferromagnetic Ising problems 
--- essentially related to the MaxCut combinatorial optimization --- on regular periodic graphs, by exploiting the intrinsic flexibility in the boundary conditions 
of a reduced spin problem. 
We prove rigorous bounds to the relative error $\eres_{\Ptrot}$ of the QAOA optimal solution on a circuit of depth $\Ptrot$. 
%
The technique, which can be naturally extended to higher-dimensional problems, is here applied to the 
translationally invariant quantum Ising chain, which is analytically tractable.
In one dimension, there is clear numerical evidence~\cite{Mbeng_optimal-dQA_arXiv2019} 
--- obtained by mapping the spin problem to free fermions through the Jordan-Wigner transformation ---
that the variational bound is precisely saturated, provided $2\Ptrot  < N$, where $N$ is the size of the Ising chain.
When $2\Ptrot=N$ the system is controllable~\cite{Dalessandro2007, Brif_NewJPhys2010}, namely the residual energy drops to $0$, and for $2\Ptrot>N$ 
the manifold of optimal solutions becomes a continuum. 

Second, we illustrate and elaborate on the link --- reported in Ref.~\onlinecite{Mbeng_optimal-dQA_arXiv2019} --- between optimal Quantum Control, 
and the adiabatic dynamics behind QA, or more precisely here {\em digitized}-QA~\cite{Martinis_Nat16}.
Indeed, as explicitly demonstrated in Ref.~\onlinecite{Mbeng_optimal-dQA_arXiv2019} for the Ising chain problem, among the large number of QAOA optimal solutions,
one can iteratively single-out a {\em smooth regular} solution which can be regarded as the optimal digitized-QA schedule, satisfying all the 
expected reasonable requirements for adiabaticity in a digitized context, without any need for spectral information on the Hamiltonian.
This comes with the extra bonus that the construction of such solution is computationally less expensive than searching for unstructured QAOA solutions. 
Here, we explicitly demonstrate the digitized {\em adiabatic} nature of such a regular solution by studying the Shannon entropy of the 
digitized evolution state on the basis of the instantaneous eigenvectors. 

The paper is organized as follows. 
Section \ref{sec:intro_QA_dQA_OC} introduces the readers to the different methods we will be dealing with, starting from QA, 
proceeding with the idea of digitalization, and then to QAOA and optimal Quantum Control.   
Section \ref{sec:qaoa} sets the MaxCut problem in general, and introduces a general technique, involving boundary conditions in a reduced chain,
which allows us to prove the residual energy bound. 
In Sec.~\ref{sec:QAOA_dQA}, following Ref.~\onlinecite{Mbeng_optimal-dQA_arXiv2019}, we exploit a Jordan-Wigner transformation to cast the multivariate minimization 
into a much simpler geometric problem, and illustrate the procedure~\cite{Mbeng_optimal-dQA_arXiv2019} to construct optimal digitized-QA adiabatic solutions, 
comparing it with other QA strategies.   
Finally, Sec.~\ref{sec:conclusions} contains a summary of our results, a discussion of relevant points, and an outlook on open questions.

\section{QA, Digitalization and Optimal Control} \label{sec:intro_QA_dQA_OC}
In this section we describe the working principle of both QA and QAOA. 
In particular we show how, in the context of schedule optimization, the QAOA algorithm emerges naturally from a digitalization of QA protocols.

\subsection{From Classical to Quantum Optimization}
Applying QA or QAOA to solve a classical optimization problem, requires a common preliminary step. 
We need to set up a quantum Hamiltonian $\Ham_z$ that encodes the original optimization problem.

Let us consider the usual setting of a cost function $C(z_1,\cdots,z_N)$ of $N$ binary variables $z_j=\pm 1$ which we want to minimize (or maximize). 
The standard strategy in quantum optimization algorithms~\cite{Finnila_CPL94,Kadowaki_PRE98,Brooke_SCI99,Santoro_SCI02,Farhi_SCI01,Farhi_arXiv2014} 
is to map $z_j \to \PauliSigma^z_j$ and regard the cost-function as a quantum Hamiltonian 
$C(z_1,\cdots,x_N)\to \Ham_z(\PauliSigma^z_1,\cdots,\PauliSigma^z_N)$, which is then minimized by employing quantum resources. 
This approach is quite general, as many optimization problems can be reformulated in this framework~\cite{Lucas2014}. 

Here, following Ref.~\onlinecite{Farhi_arXiv2014}, we will illustrate it for the MaxCut problem \cite{Boros_AnnOpRes1991}.
Given a graph, i.e., a set of vertices or nodes ${\mathcal G}=\{j\}$ connected by certain edges ${\mathcal E}=\{e\}$, 
the MaxCut problem consists in finding the largest number of edges that need to be cut when partitioning the graph into two independent parts. 
By assigning a label $z_j=+1$ and $z_j=-1$ to the nodes of the two independent parts, the objective cost function to be maximized is 
$C_{\scriptstyle{\mathrm{MaxCut}}}(\boldsymbol{z}) = \sum_{\left\langle i,j\right\rangle \in \mathcal{E} }(1-z_iz_j)/2$, where the sum runs on all the edges 
$\left\langle i,j\right\rangle = e\in {\mathcal E}$.
This amounts, in the quantum language, to searching for the minimum of
\begin{equation} \label{eqn:Hz_classical}
\Ham_z = \sum_{\left\langle i,j\right\rangle  \in \mathcal{E} }(\PauliSigma^z_i \PauliSigma^z_j-1)
\end{equation}
which we denote by $E_{\min}$. Notice that, having omitted here the factor $1/2$, $|E_{\min}|$ is {\em twice} the maximum number of cut edges. 
The maximum eigenvalue of $\Ham_z$ is clearly $0$, $E_{\max}=0$. 

\subsection{Continuous and digitized quantum annealing}
According to standard continuous-time QA \cite{Kadowaki_PRE98,Brooke_SCI99,Santoro_SCI02,Farhi_SCI01},
the cost Hamiltonian $\Ham_z$ has to be supplemented by a driving term, usually --- but not necessarily \cite{Seoane2012} --- taken to be of the simple form 
$\Ham_x=-\sum_{j=1}^N \PauliSigma^x_j$.
In the simplest setting, one would then write a QA Hamiltonian of the form:
\begin{equation} \label{eqn:Hs}
\Ham(s) = s \, \Ham_z + (1-s) \, \Ham_x \;.
\end{equation}
The parameter $s$ is then varied in time, defining a schedule $s(t)$ interpolating between $s(0)=0$ and $s(\Tann)=1$, where
$\Tann$ is the total annealing time. 
In its simplest form, QA is often associated to a {\em linear schedule} $s(t)=t/\Tann$, but this restriction can be in principle removed, as 
one might optimize the schedule $s(t)$ appropriately~\cite{Roland_PRA2002, Polkovnikov_PRL2008, Zanardi_PRL2009, Caneva_PRL2009, Caneva_PRA2011}.
Given a schedule $s(t)$, the QA algorithm works as follows: first we initialize the system
in the ground state of $\Ham_x$, 
\begin{equation}
|\psi_0\rangle=|+\rangle^{\otimes N} = \left( \frac{ |\!\uparrow\rangle + |\!\downarrow\rangle}{\sqrt{2}} \right)^{\otimes N} \;.
\end{equation}
Then we let it evolve for a total annealing time $\tau$ under the action of $\Ham(s(t))$. 
We can describe the QA dynamics with a unitary quantum Schr\"odinger evolution~\cite{Messiah:book}: the system's state at any given $t$ is 
$|\psi(t)\rangle = \Uevol_{\QA}(t,0) | \psi_0 \rangle$, where
\begin{equation} \label{eqn:U_timeordered}
\Uevol_{\QA}(t,0) = \Texp \left(-\frac{i}{\hbar} \int_0^t \! \ud t' \, \Ham(s(t')) \right) \;,
\end{equation} 
where $\Texp$ denotes the time-ordered exponential. 

\begin{figure}
    \includegraphics[width=\columnwidth]{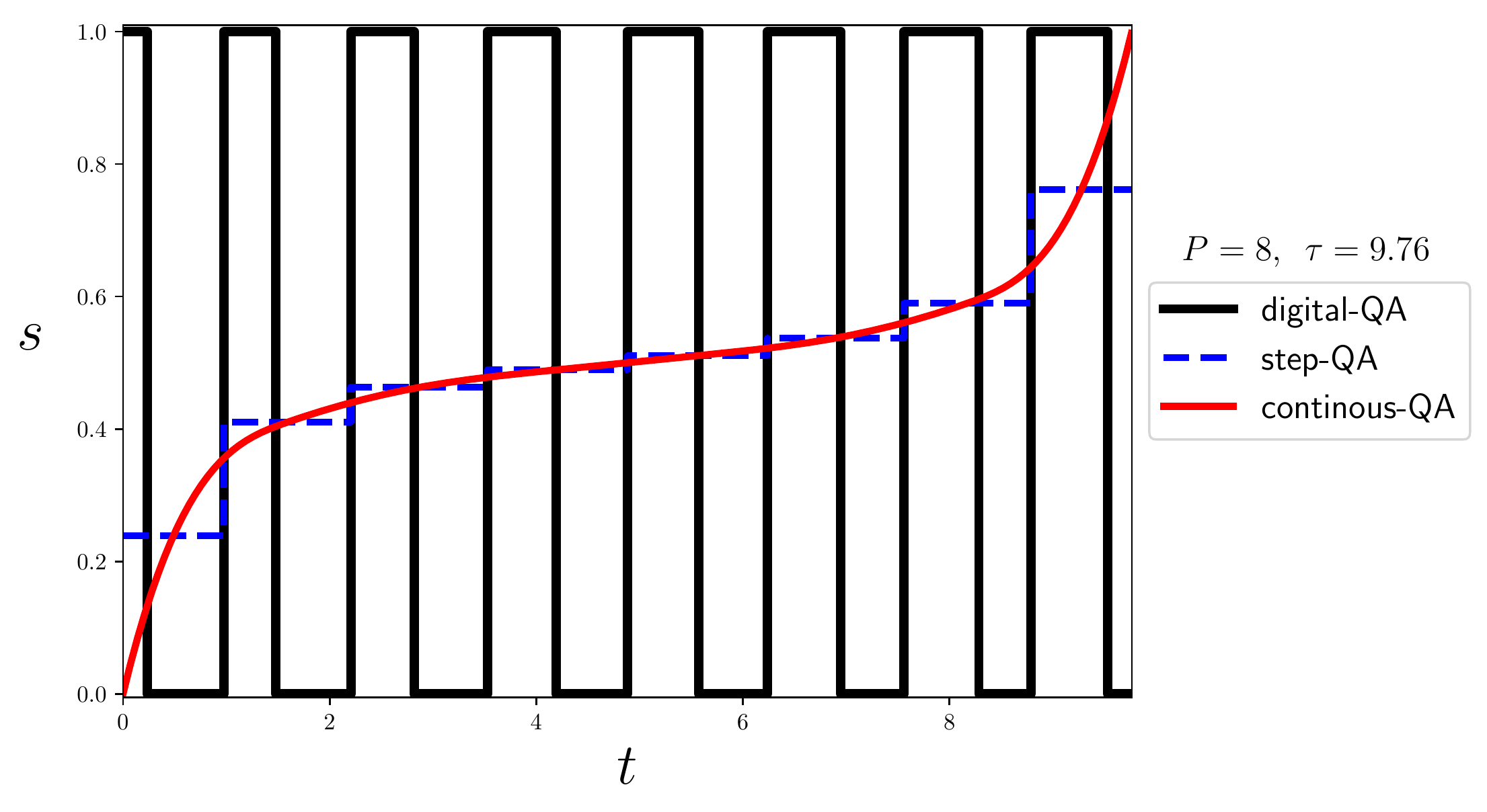}
    \caption{A smooth continuous-time QA $s(t)$, with the associated step-QA and digitized-QA schedules. Here $\Ptrot=8$.
    Notice that the time-intervals $\Delta t_m$ are not identical. 
    The procedure by which this figure is generated is explained in Sec.~\ref{sec:from_QAOA_to_QA}. }
    \label{fig:bang-bang}
\end{figure}
%
In many situations it is meaningful, in some cases necessary, to approximate the schedule $s(t)$ by a step function attaining $\Ptrot$ values $s_1,\cdots, s_\Ptrot$, 
with $s_m\in (0,1]$, corresponding to evolution times $\Delta t_m$, with $m=1,\cdots, \Ptrot$ such that $\sum_{m=1}^\Ptrot \Delta t_m = \Tann$. 
Figure~\ref{fig:bang-bang} is a sketch of such a step-discretization starting from a smooth $s(t)$ 
--- which we might refer to as a {\em step-QA} scheme --- but the discussion below applies to {\em any} step function. 
The evolution operator $\Uevol(\Tann,0)$ is then given by:
\begin{equation} \label{eqn:U_step}
\Uevol_{\QA}(\Tann,0) \Longrightarrow \Uevol_{\sgdigital} = \Tprod{\Ptrot}_{m=1} \ee^{-\frac{i}{\hbar} \Ham(s_m) \Delta t_m} \;.
\end{equation}
where the arrow $\leftarrow$ denotes a time-ordered product.
 
A further digitalization step would be to perform a {\em Trotter splitting} of the term $\ee^{-\frac{i}{\hbar} \Ham(s_m) \Delta t_m}$. 
For instance, the lowest-order Trotter splitting
\begin{equation} \label{eqn:Trotter_lowest}
\ee^{-\frac{i}{\hbar} \Ham(s_m) \Delta t_m}  \simeq \ee^{-i\beta_m \Ham_x} \ee^{-i \gamma_m \Ham_z} + O((\Delta t_m)^2) \; 
\end{equation}
with
\begin{equation} \label{eqn:gamma_beta_m}
\left\{ 
\begin{array}{l}
\gamma_m = \displaystyle s_m \frac{\Delta t_m}{\hbar} \vspace{2mm}\\
\beta_m = \displaystyle (1-s_m) \frac{\Delta t_m}{\hbar}
\end{array}
\right. 
\end{equation}
would lead to an approximate evolution operator of the form:
\begin{equation} \label{eqn:Udigital}
\Uevol_{\QA}(\Tann,0) \approx \Uevol_{\mgdigital}(\bgamma,\bbeta) = \Uevol(\gamma_{\Ptrot},\beta_{\Ptrot}) \cdots \Uevol(\gamma_1,\beta_1) \;, 
\end{equation}
with
\begin{equation} \label{eqn:Udigital_bis}
\Uevol(\gamma_m,\beta_m) \equiv \Uevol_m = \ee^{-i\beta_m \Ham_x} \ee^{-i \gamma_m \Ham_z} \;.
\end{equation}
The parameters $\gamma_m$ and $\beta_m$ are such that:
\begin{equation} \label{eqn:sum_rule}
\sum_{m=1}^\Ptrot (\gamma_m + \beta_m) = \frac{\Tann}{\hbar} \;.
\end{equation} 
If the original scheme was a QA based on a smooth $s(t)$, then we might call this a {\em digitized-QA} (dQA)~\cite{Martinis_Nat16}.  
A symmetric, or any higher order, Trotter splitting would lead to modified expressions for the parameters 
$\bgamma=(\gamma_1, \cdots, \gamma_\Ptrot)$ and $\bbeta=(\beta_1, \cdots, \beta_\Ptrot)$, 
with identical sum rule \eqref{eqn:sum_rule}. 

\subsection{QAOA and optimal quantum control}
Eq.~\eqref{eqn:Udigital} describes a time evolution obtained by alternating the two Hamiltonians $\Ham_z$ and $\Ham_x$. 
This is the starting point of the Quantum Approximate Optimization Algorithm (QAOA) introduced in Ref.~\onlinecite{Farhi_arXiv2014}.

In the QAOA, we use a digital device that alternates the application of $\Ham_z$ and $\Ham_x$ to prepare the quantum variational state
\begin{equation} \label{eqn:psi_qaoa}
|\psi_\Ptrot(\bgamma,\bbeta)\rangle   
= \Uevol(\gamma_{\Ptrot},\beta_{\Ptrot}) \cdots \Uevol(\gamma_{1},\beta_{1}) | \psi_0\rangle \;,
\end{equation}
which depends on the $2\Ptrot$ free parameters $(\bgamma,\bbeta)$. Then, we compute the expectation value of the cost function 
\begin{equation}
E_\Ptrot(\bgamma,\bbeta) = \langle \psi_\Ptrot(\bgamma,\bbeta) |  \Ham_z | \psi_\Ptrot(\bgamma,\bbeta) \rangle \;,  
\end{equation}
by repeated measurements in the computational basis. 
%
The global minimum $(\bgamma^{*},\bbeta^{*})$ of the variational parameters determines a correspondingly optimal energy
$E_\Ptrot^{\opt}=E_\Ptrot(\bgamma^{*},\bbeta^{*})$ which is, by construction, a monotonically decreasing function of $\Ptrot$.
%
%
An approximate solution of the classical problem can be obtained by repeated measurements on the state $\ket{\psi_\Ptrot(\bgamma^*, \bbeta^*)}$.
The total evolution ``time'' $\Tann$, however, is no longer fixed, but rather related to the optimal 
parameters $(\bgamma^*, \bbeta^*)$ by the sum rule in Eq.~\ref{eqn:sum_rule}.
For Boolean Satisfiability~\cite{Garey_Johnson:book} problems, where $\Uevol_{\mgdigital}(\bgamma,\bbeta)$ turns out to be periodic in the variational parameters, 
one might still show that $\tau<2\pi \Ptrot$.
Remarkably, the QAOA approach has been shown to be computationally universal~\cite{Lloyd_arXiv2018}, although this fact does not guarantee, by
itself, efficiency or speedup~\cite{Hastings_arXiv2019}. 


An interesting recent result~\cite{Yang_PRX2017} concerns optimal Quantum Control~\cite{Dalessandro2007, Brif_NewJPhys2010}. 
Indeed, suppose that the total evolution time $\Tann$ is fixed, and one asks for a the optimal schedule $s(t)$ with values bounded in the interval $[0,1]$ 
without any continuity or monotonicity requirement. 
Then, as shown by Yang \textit{et al.}~\cite{Yang_PRX2017}, an application of Pontryagin's principle shows that
the optimal schedule has to be of the so-called {\em bang-bang} form, with $s(t)$ having a square-wave form between the
two extremal values $1$ and $0$, as sketched in Fig.~\ref{fig:bang-bang}.
Denoting by $\gamma_m$ and $\beta_m$ the ``time-lags'' spent in the $m$-th intervals with $s=1$ and $s=0$, respectively, we recover once again 
the form given in Eq.~\eqref{eqn:Udigital} or \eqref{eqn:psi_qaoa} where, however, the total number of recursions $\Ptrot$ is no longer fixed. 

These considerations point towards a crucial point that we elucidate in our paper. 
While, by definition, any digitized-QA schedule is of the bang-bang form, it is {\it a priori} not obvious {\em if and when} an optimal control bang-bang solution can be constructed 
which is also {\em adiabatic}. 
Here and in Ref.~\onlinecite{Mbeng_optimal-dQA_arXiv2019} we shed light on the last point, by explicitly constructing 
an adiabatic optimal schedule for the translationally invariant quantum Ising chain case, and arguing that it is indeed computationally convenient to do so, 
rather than searching for an arbitrary unstructured optimal solution. 
%

\section{Residual energy bound for MaxCut on $2$-regular graphs} \label{sec:qaoa}
%
Let us turn our attention to the performance of the QAOA algorithm on MaxCut problems. 

The QAOA aims at finding a variational state that minimizes 
$E_\Ptrot(\bgamma, \bbeta) = \bra{{\psi_\Ptrot(\bgamma,\bbeta)} }\Ham_z\ket{\psi_\Ptrot(\bgamma, \bbeta)}$, the expectation value of the cost-function Hamiltonian.
%
%
To quantify how well a given variational state $\ket{\psi_\Ptrot(\bgamma,\bbeta)}$ approximates the solution of the optimization problem, we introduce the residual 
energy~\cite{Santoro_SCI02}
\begin{equation} \label{eqn:e_res} 
\eres_{\Ptrot}(\bgamma, \bbeta) =  \frac{E_\Ptrot(\bgamma, \bbeta)-E_{\min}}{E_{\max}-E_{\min}}\;.
\end{equation}
With the given normalization, $\eres_\Ptrot(\bgamma, \bbeta)=0$ if and only if $\ket{\psi_\Ptrot(\bgamma,\bbeta)}$ is a solution to the optimization problem encoded by $\Ham_z$. 
For a generic state, $\eres_{\Ptrot}(\bgamma, \bbeta)\in[0,1]$ represents the relative approximation error.
This quantity is related to the approximation ratio $r_\Ptrot$ considered in the context of QAOA~\cite{Farhi_arXiv2014, Wang_PRA2018, Lukin_arXiv2018} 
by the simple relation $r_{\Ptrot}=1-\eres_{\Ptrot}$.

By evaluating $\eres_{\Ptrot}$ (or $r_{\Ptrot}$) on the output of the QAOA we get the best residual energy $\left(\eres_{\Ptrot}\right)^{*}=\eres_{\Ptrot}(\bgamma^*, \bbeta^*)$,
or $r^*_\Ptrot = r_\Ptrot(\bgamma^*, \bbeta^*)$, a natural figure of merit for the performance of the algorithm~\cite{Farhi_arXiv2014}.
Clearly, increasing the number of variables improves the QAOA's performance, hence $\left(\eres_{\Ptrot+1}\right)^{*} \le  \left(\eres_{\Ptrot}\right)^{*}$
or, equivalently, $r^*_{\Ptrot+1}\geq r^*_\Ptrot$. 
Moreover, if the Hamiltonian $\Ham(s) = s \, \Ham_z + (1-s) \, \Ham_x $ has a finite minimum gap in the interval $s\in [0,1]$, an arbitrarily good solution can be obtained by 
Trotterization of an adiabatic evolution, so that $\left(\eres_{\Ptrot}\right)^{*}\to 0$ and $r^*_\Ptrot \to 1$ for $\Ptrot\to\infty$ ~\cite{Farhi_arXiv2014}. 

Following Ref.~\onlinecite{Farhi_arXiv2014, Wang_PRA2018, Lukin_arXiv2018, Crooks_arXiv2018}, we consider the MaxCut problem restricted to specific classes of graphs. 
The simplest cases are connected $\kconn$-regular graphs, where the connectivity (or degree) $\kconn$, i.e., 
the number of edges originating from each node, is fixed and identical for all nodes. 

A $2$-regular graph is a closed ring. In this case, the MaxCut problem --- sometimes referred to as {\em ring of disagrees} \cite{Farhi_arXiv2014} --- 
is equivalent to finding the classical ground state of an anti-ferromagnetic Ising model on a chain, a computationally ``easy'' problem which, physically, 
shows no frustration (for $N$ even).
Generic $\kconn$-regular graphs with $\kconn\ge 3$ are frustrated and solving the associated Max-Cut problem is hard.   
For $3$-regular graphs, Ref.~\onlinecite{Farhi_arXiv2014} has shown that the {\em worst case} value of the approximation ratio $r_\Ptrot$ is bound, for $\Ptrot=1$, to be
$r_{\Ptrot=1}^*\ge 0.6924$. 

Here we will show that for the $2$-regular case, or equivalently the antiferromagnetic Ising ring, one can prove a lower bound for the residual 
energy $\eres_{\Ptrot}$ or, equivalently an upper bound for $r_\Ptrot$:
\begin{equation} \label{eqn:eres_final_main}
\eres_{\Ptrot} \ge 
\left\{ \begin{array}{ll} \displaystyle \frac{1}{2 \Ptrot + 2} & \mbox{for} \hspace{2mm} 2\Ptrot < N \vspace{3mm} \\
0  & \mbox{for} \hspace{2mm} 2\Ptrot \ge N
\end{array} \right. \;.
\end{equation}


 
\begin{figure}
    \includegraphics[width=\columnwidth]{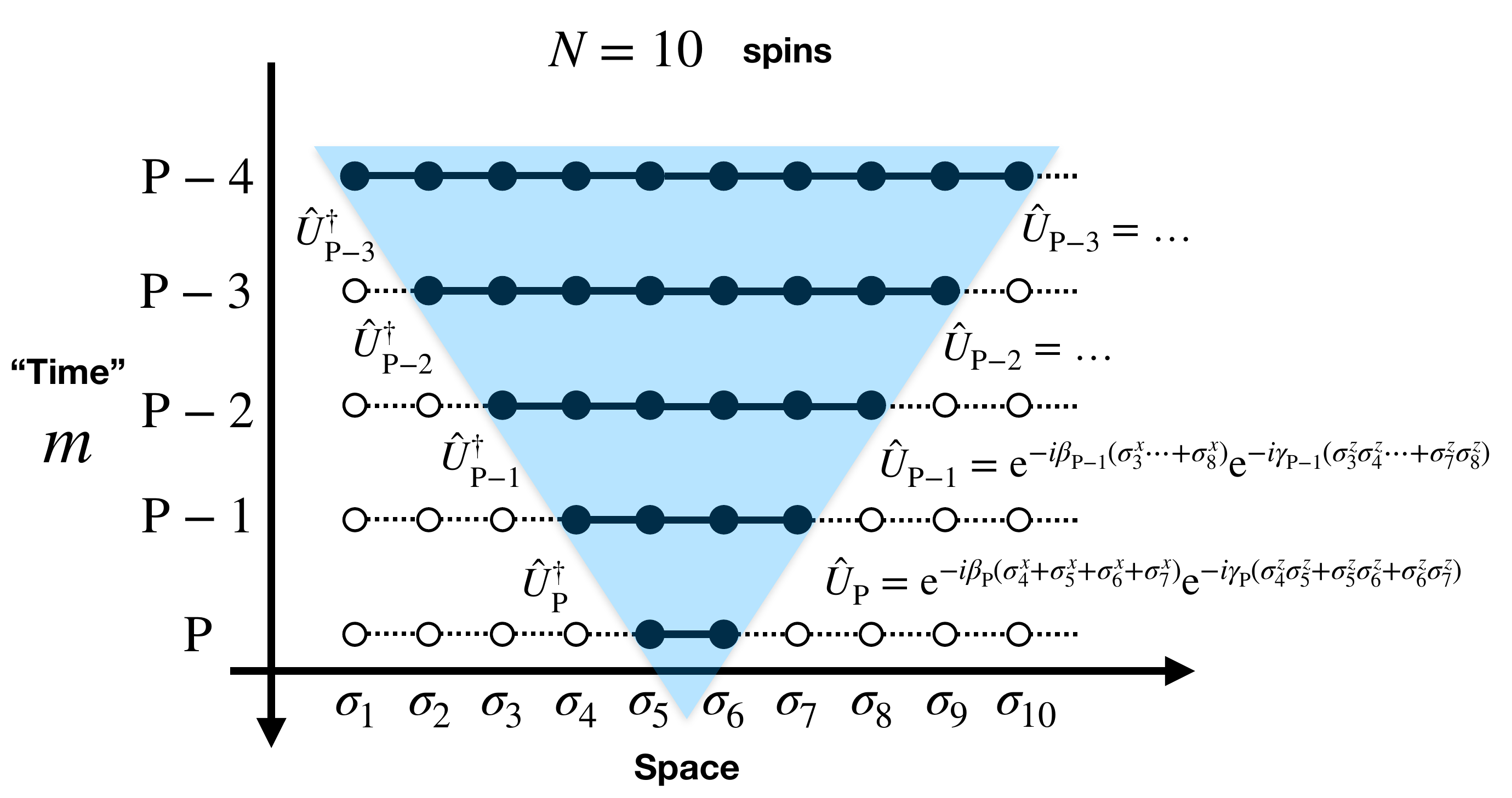}
    \caption{A sketch of how the successive application of digitized evolution results in an operator spreading
    that justifies the use of a reduced spin chain. 
    Notice the systematic absence of boundary terms in the reduced chain Hamiltonian.}
    \label{fig:light_cone}
\end{figure}

To prove this result, we start by considering the translational invariance of $\Ham_z$, $\Ham_x$ and of the initial state $|\psi_0\rangle$. 
Indeed, because of translational invariance we can write the residual energy in Eq.~\eqref{eqn:e_res} as:
\begin{equation}
\eres_{\Ptrot}(\bgamma,\bbeta) = 
\bra{\psi_\Ptrot(\bgamma,\bbeta)}\frac{\PauliSigma^z_{j_s} \PauliSigma^z_{j_s+1} + 1}{2}\ket{\psi_\Ptrot(\bgamma,\bbeta)} \;, 
\end{equation}
where $j_s$ is any site of the chain, for instance the central site $j_s=N/2$. (We restrict ourself to even $N$ so that the chain is unfrustrated.)
As demonstrated in Ref.~\onlinecite{Farhi_arXiv2014}, the application of the digitized unitary operator 
$\Uevol_{\mgdigital}(\bgamma,\bbeta)=\Uevol_{\Ptrot} \cdots \Uevol_1$ --- where $\Uevol_m = \Uevol(\gamma_m,\beta_m)$ --- 
to link operator $\PauliSigma^z_{j_s} \PauliSigma^z_{j_s+1}$
\begin{equation}
\Uevol_{1}^{\dagger} \cdots \Uevol_{\Ptrot}^{\dagger} \, \PauliSigma^z_{j_s} \PauliSigma^z_{j_s+1} \, \Uevol_{\Ptrot} \cdots \Uevol_1
\end{equation}
involves only spins which have a distance at most $\Ptrot$ from the link $(j_s,j_s+1)$. 
Such an operator spreading is sketched in Fig.~\ref{fig:light_cone}, where a certain similarity with the light-cone idea emerges~\cite{Chuang_arXiv2019}. 
Considering for instance the central link $j_s=\frac{N}{2}$, if $(2\Ptrot+2)<N$, this leads to a reduced spin chain with
$\Nred=(2\Ptrot+2)$ sites, $j=j_s-\Ptrot, \cdots, j_s+\Ptrot+1$, while if $(2\Ptrot+2)\ge N$, all sites are involved, $\Nred=N$.  
%
Without loss of generality, we can always re-number the sites belonging to the reduced spin chains from $1$ to $\Nred$, and assume
that the link we are ``measuring'' is the central one, $j_s=\Nred/2$.

Notice that for $(2\Ptrot+2)=\Nred\le N$, in the spirit of the spin reduction explained above, 
any boundary term is {\em absent} in the reduced chain Hamiltonian. 
Hence, we are free to add an {\em arbitrary} boundary link term $J_b \, \PauliSigma^z_{\Nred} \PauliSigma^z_{1}$.
We restrict our choice here to $J_b=\pm 1$. 
For $J_b=+1$ the reduced spin chain has periodic boundary conditions (PBC), hence recovering full translational invariance. 
For $J_b=-1$ we have anti-periodic boundary conditions (ABC), but an effective translation operator involving a spin-flip can still be 
introduced (see App.~\ref{app:ABC_to_PBC}).
Our claim now is that we can work with a reduced spin chain Hamiltonian of the form:
\begin{equation} \label{eqn:Hred_z}
\Hred_z^{\smallpm} = \sum_{j=1}^{\Nred-1}\left(\PauliSigma^z_j \PauliSigma^z_{j+1}-1\right) + \left( \pm \, \PauliSigma^z_{\Nred} \PauliSigma^z_{1} -1\right) \;,
\end{equation}
\noindent
while keeping the transverse term unmodified
\begin{equation}\label{eqn:Hred_x}
\Hred_x = -\sum_{j=1}^{\Nred} \PauliSigma^x_j \;,
\end{equation}
and show that this would not modify the expectation value we need, i.e.,
\begin{equation} \label{eqn:eres_invariant}
\eres_{\Ptrot}(\bgamma,\bbeta)
= \bra{ \widetilde{\psi}_\Ptrot(\bgamma,\bbeta) } \frac{\PauliSigma^z_{j_s} \PauliSigma^z_{j_s+1}+1}{2}  
   \ket{ \widetilde{\psi}_\Ptrot(\bgamma,\bbeta) }
\end{equation}
where
\begin{equation}\label{eqn:psi_tilde_qaoa}
 \ket{ \widetilde{\psi}_\Ptrot(\bgamma,\bbeta) } = \Tprod{\Ptrot}_{m=1} \ee^{-i \beta_m \Hred_x} \ee^{-i \gamma_m \Hred_z^{\smallpm}} \; | \widetilde{\psi}_0\rangle \;,
\end{equation}
with $\ket{ \widetilde{\psi}_0 } = |+\rangle^{\otimes \Nred}$. 
Such extra freedom in the boundary conditions will be used shortly to derive a bound for the residual energy. 
We notice that when $2\Ptrot+2>N$ the reduced spin chain coincides with the full chain, $\Nred=N$, and there is no freedom whatsoever: we must
use PBC.

Returning to the case $2\Ptrot+2=\Nred \le N$, the next crucial step is to show that:
\begin{eqnarray} \label{eqn:antiperiodic_average_main}
\bra{\widetilde{\psi}_\Ptrot} \left(\PauliSigma^z_{j_s} \PauliSigma^z_{j_s+1}-1\right) \ket{\widetilde{\psi}_{\Ptrot}} &=&
\frac{1}{\Nred} \bra{\widetilde{\psi}_{\Ptrot}} \Hred_z^{\smallpm} \ket{\widetilde{\psi}_{\Ptrot}}  \;.
\end{eqnarray}
For the PBC case, this is a trivial consequence of translational invariance. 
For the ABC case, one needs to show that a modified translation operator can be introduced, by incorporating a spin-flip at site $1$, which
does the job: this is shown explicitly in App.~\ref{app:ABC_to_PBC}.
Using Eqs.~\eqref{eqn:antiperiodic_average_main} and \eqref{eqn:eres_invariant} we immediately conclude that 
%
\begin{eqnarray} \label{eqn:eres_final_ABC}
\eres_{\Ptrot}(\bgamma,\bbeta)
&=& \bra{ \widetilde{\psi}_\Ptrot(\bgamma,\bbeta) } \left( \frac{ \Hred^{\smallpm}_z}{2(2\Ptrot+2)} +1 \right)  
 \ket{ \widetilde{\psi}_\Ptrot(\bgamma,\bbeta) } \nonumber \\
&\ge& \left( \frac{E^{\smallpm}_{\gs}}{2(2\Ptrot+2)}  +1 \right) \;,
\end{eqnarray}
where $E^{\smallpm}_{\gs}$ is the ground state energy of $ \Hred^{\smallpm}_z$ and the inequality follows from the standard variational principle. 
Now observe that $E^{\smallp}_{\gs}=-2\Nred = -2(2\Ptrot+2)$. 
Hence, if PBC are used in the reduced spin chain, the bound is a trivial inequality $\eres_{\Ptrot}(\bgamma,\bbeta)\ge 0$. 
The inequality, however, becomes non-trivial if ABC are used, since $E^{\smallm}_{\gs}=-4\Ptrot-2$ due to the frustrating boundary term $J_b=-1$:
\begin{equation} \label{eqn:eres_final}
\eres_{\Ptrot}(\bgamma,\bbeta)
\ge \left( \frac{E^{\smallm}_{\gs}}{2(2\Ptrot+2)}  +1 \right)  = \frac{1}{2\Ptrot +2} \;,
\end{equation}
This establishes the promised bounds in Eq.~\eqref{eqn:eres_final_main}. 

As a check, observe that the computation of the optimal $r_{\Ptrot=1}^*$ carried out in Ref.~\onlinecite{Wang_PRA2018} (Eq.~(16))  
and valid for triangle-free $\kconn$-regular graphs, when translated into our notation would imply that:
\begin{equation}\label{eqn:eres_P1}
    \eres_{\Ptrot=1} \ge \frac{1}{2} - \frac{\left(1-\frac{1}{\kconn}\right)^{\frac{\kconn-1}{2}}}{2\sqrt{\kconn}} \;.
\end{equation}
This shows that the optimal result in Eq.~\eqref{eqn:eres_P1}, when specialized to $\kconn=2$, coincides with Eq.~\eqref{eqn:eres_final}.
Later on, see Fig.~\ref{fig:eres_bound_vs_P} and accompanying discussion, we will explicitly demonstrate numerically the tightness of the bound in 
Eq.~\eqref{eqn:eres_final} also for larger $\Ptrot$. 

The derivation of the bound in Eq.~\eqref{eqn:eres_final_main} given here relies on the locality and translational invariance of the problem. 
It can therefore be generalized to systems in higher dimensions, as explicitly shown in Ref.~\onlinecite{Mbeng_PhDThesis2019}.  


\section{Jordan-Wigner results: attaining the variational bound} \label{sec:QAOA_dQA}
For the $2$-regular graph case, i.e., the antiferromagnetic Ising chain case, a lot more can be said.
First of all, as discussed in Ref.~\onlinecite{Wang_PRA2018}, one can take advantage of the Jordan-Wigner transformation \cite{JordanWigner_ZPhys1928}
to map the problem into a free-fermion one. In particular, one can show that the system is equivalent to a set of independent two-level systems. 

Let us recall, for the reader's convenience, that if $2\Ptrot+2>N$, then we are dealing with a standard Ising chains of length $\Nred=N$ with
the original (PBC) boundary conditions; otherwise, if $2\Ptrot+2=\Nred\le N$, the reduced spin chain discussed in the previous section can be
taken to have an {\em arbitrary} boundary term. As discussed previously, using ABC leads to the bound we have derived.
Hence, $\Nred=\min(N,2\Ptrot+2)$ is the effective chain length, and we will set the boundary condition appropriately. 
 
Using a Jordan-Wigner transformation (see Ref.~\onlinecite{Mbeng_optimal-dQA_arXiv2019} and accompanying Supplementary Information for details), 
the many-body problem can be decomposed into a collection of independent two-level systems, or pseudo-spins. 
The set of wave-vectors labelling the pseudo-spins depends on the boundary conditions assumed for the reduced spin. 
For PBC and ABC, the sets of wave-vectors are respectively $\calK_{\PBC}=\{ \frac{\pi}{\Nred}, \frac{3 \pi}{\Nred} \cdots,\frac{ (\Nred-1)\pi }{\Nred} \}$ and 
$\calK_{\ABC}=\{ \frac{2\pi}{\Nred}, \frac{4\pi}{\Nred}, \cdots, \frac{(\Nred-2)\pi}{\Nred} \}$.
The total residual energy $\eres_{\Ptrot}(\bgamma, \bbeta)$, is a sum of the contributions arising from each two-level system.  
For $2\Ptrot < N$ (where ABC on the reduced chain are allowed) the result is:
\begin{equation}  \label{eqn:eres_abc}
    \eres_{\Ptrot}(\bgamma, \bbeta) \stackrel{\scriptscriptstyle 2\Ptrot< N}{=}  
    \frac{1}{2\Ptrot+2} + \frac{1}{2\Ptrot+2}{\displaystyle \sum_{k}^{\calK_{\ABC}}} \epsilon_{k}(\bgamma, \bbeta) \;,
\end{equation}
while for $2\Ptrot\ge N$ (where only PBC are allowed) we get:
\begin{equation}  \label{eqn:eres_pbc}
    \eres_{\Ptrot}(\bgamma, \bbeta) \stackrel{\scriptscriptstyle 2\Ptrot\ge N}{=}  \frac{1}{N}{\displaystyle \sum_{k}^{\calK_{\PBC}}} \epsilon_{k}(\bgamma, \bbeta) \;.
\end{equation}
Here, the non negative contribution arising from the two-level-system of wave-vector $k$ is
\begin{equation} \label{eqn:eresk_geometrical_def}
    \epsilon_k(\bgamma, \bbeta)= 1-  \versorb_k^T \left( \Tprod{\Ptrot}_{m=1}  \Rrot_{\versorz}(4 \beta_m )\Rrot_{\bvec_k}(4 \gamma_m ) \right) 
    \versorz \in [0,1]\; ,
\end{equation}
which is given in terms of $3\times 3$ rotation matrices $\Rrot$ around unit vectors
$\versorz=(0,0,1)^T$ and $\versorb_{k}=(-\sin k,0, \cos k)^T$ by rotation angles $4\beta_m$ and $4 \gamma_m$, respectively.

Eqs.~\eqref{eqn:eres_abc}, \eqref{eqn:eres_pbc} and \eqref{eqn:eresk_geometrical_def} are our starting points to discuss the properties of the QAOA landscape.
The first observation is that the landscape has periodicity of $\pi/2$ in each variable $\gamma_m$ and $\beta_m$~\cite{Farhi_arXiv2014}. 
Without loss of generality we can assume $\gamma_m, \beta_m \in[0,\frac{\pi}{2}]$.

A second observation emerges from the inspection of Eq.~\eqref{eqn:eres_abc}. 
For $2\Ptrot<N$ we use ABC and therefore $\Nred=2\Ptrot+2$, 
which implies that $\eres_{\Ptrot}(\bgamma,\bbeta)$ is totally independent of $N$. 
This $N$-independence is, in retrospective, a general consequence of the spin reduction behind QAOA for translational invariant models~\cite{Farhi_arXiv2014}, 
valid well beyond the Jordan-Wigner framework used to derive Eq.~\eqref{eqn:eres_abc}.
Moreover, the optimal residual energy $\eres_{\Ptrot}(\bgamma^*,\bbeta^*)$ saturates the bound in Eq.~\eqref{eqn:eres_final_main}, hence
$\eres_{\Ptrot}(\bgamma^*,\bbeta^*)=\frac{1}{2\Ptrot+2}$, provided we are able to make the 
contribution from $\sum_k \epsilon_k$ to vanish (which we can, as discussed below).   

Finally, in the App.~\ref{app:QAOA_landscape}  we show how some simple transformation 
properties of the system translate into corresponding properties for the QAOA landscape.
In particular, one can show~\cite{Wang_PRA2018} that:
\begin{eqnarray}
    \eres_{\Ptrot}(\bgamma, \bbeta) &=& \eres_{\Ptrot}(-\bgamma, -\bbeta)  \nonumber \\
    \eres_{\Ptrot}(\bgamma, \bbeta) &=& \eres_{\Ptrot}(\bbeta', \bgamma') \;,
\end{eqnarray}
where we have defined the vectors $\bbeta^\prime, \bgamma^\prime$ as $\beta'_m = \beta_{\Ptrot-m-1}$ and $\gamma'_m = \gamma_{\Ptrot-m-1}$.
In Ref.~\onlinecite{Wang_PRA2018} it was shown that the optimal values for the parameter lie in the sub-manifold $\bbeta = \bgamma^\prime$ 
for $\Ptrot\le 10$. We have confirmed this result --- which applies to the case $2\Ptrot< N$ --- for $\Ptrot\le 128$. 

\begin{figure*}
    \includegraphics[width=55mm]{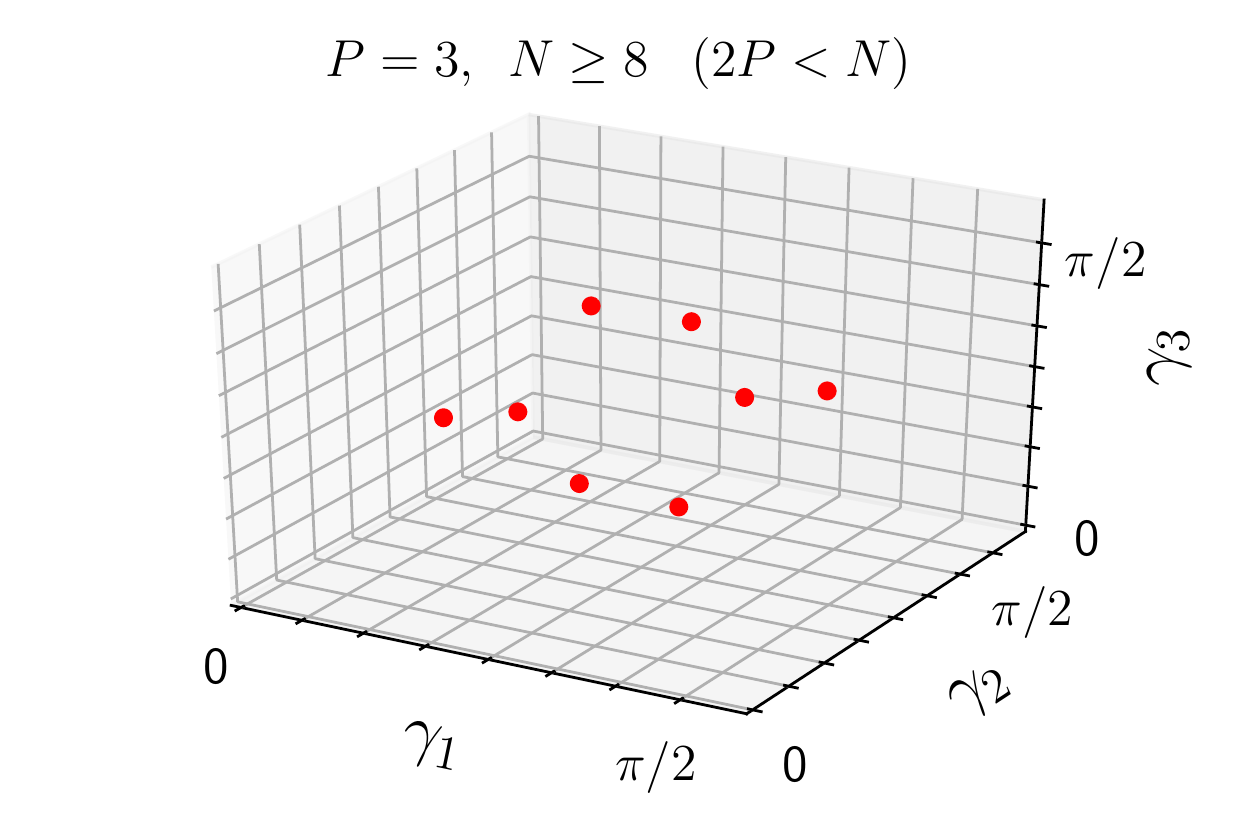}
    \includegraphics[width=55mm]{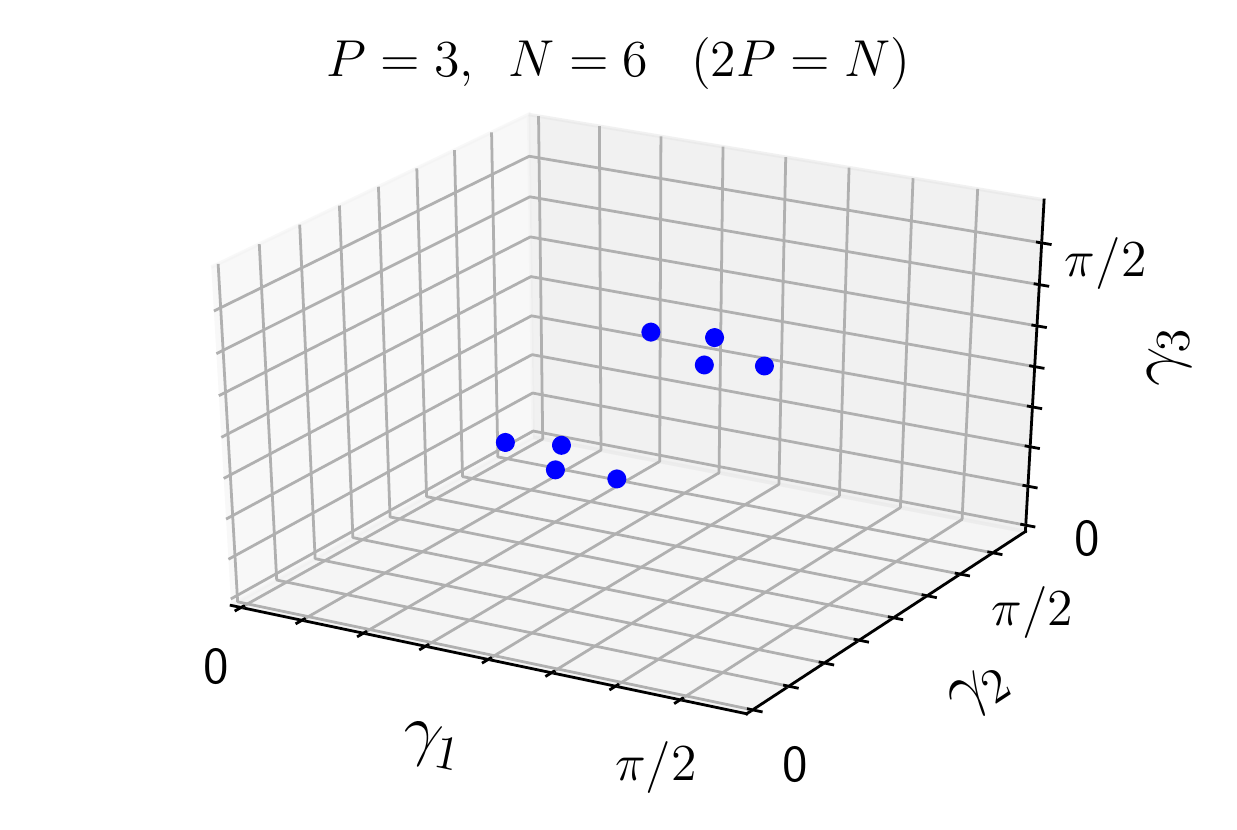}
    \includegraphics[width=55mm]{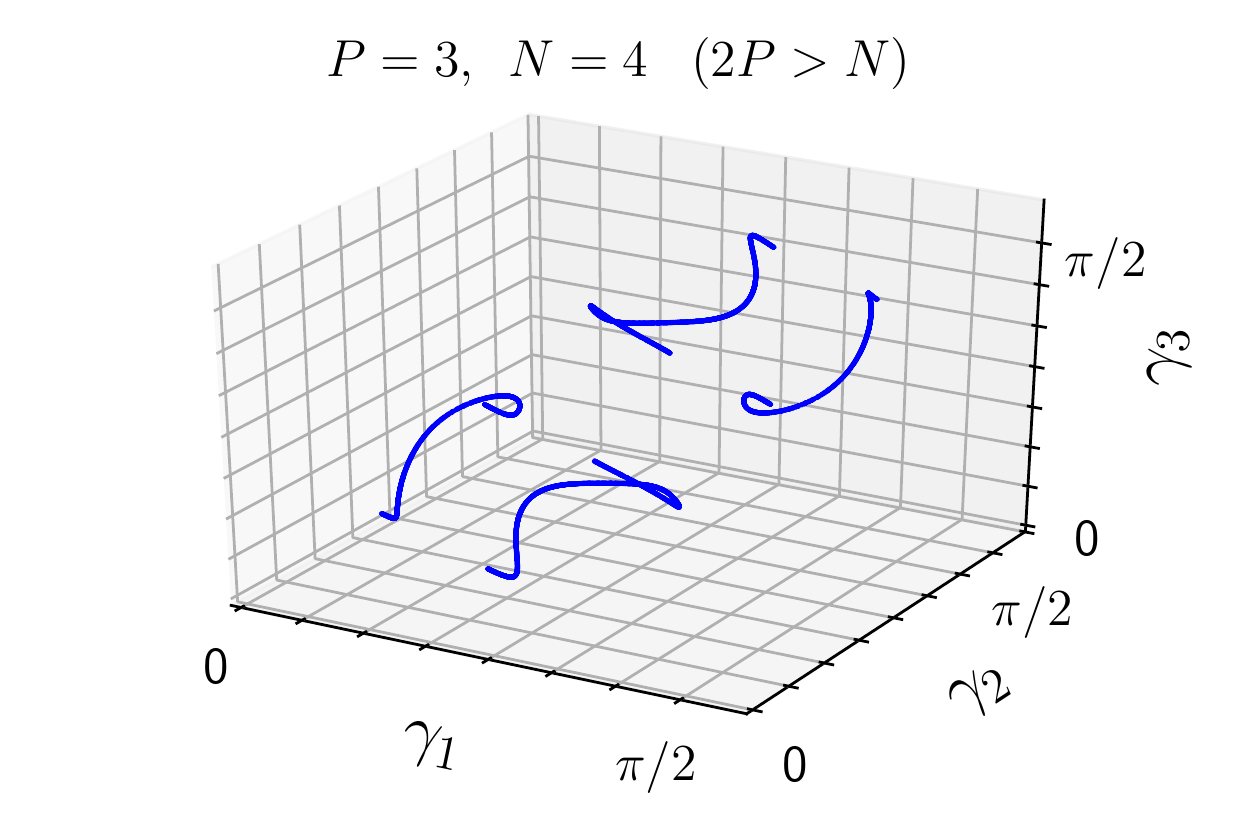}
    \caption{Visualization of the optimal solutions of QAOA for $\Ptrot=3$ in the symmetric manifold 
    $\bbeta = \bgamma^\prime$. 
    Blue circles correspond to values $\eres_{\Ptrot=3} = 0$ while red ones corresponds to the values $\eres_{\Ptrot=3} = 1/(2\Ptrot+2)=1/8$. 
    In (a), for $2\Ptrot<N$ ($N\ge 8$), the optimization problem has a finite set of solutions corresponding to strictly positive values of $\eres=1/8$. 
    In (b), for $2\Ptrot= N$ ($N=6$), there is a finite set of optimal solutions, now having $\eres=0$. 
    In (c), for $2\Ptrot>N$ (here $N=4$), the manifold of solutions attaining $\eres=0$ has dimension $2\Ptrot-N=2$: the curves shown are obtained
    by intersecting the solution surface with the symmetric manifold $\bbeta = \bgamma^\prime$.}
    \label{fig:QAOA_minima}
\end{figure*}

The function $\epsilon_{k}(\bgamma, \bbeta)$ has a simple geometrical interpretation: it contains the scalar product of $\versorb_k$ with the vector 
$\left(\Tprod{\Ptrot}_{m=1}\Rrot_{\versorz}(4 \beta_m )\Rrot_{\bvec_k}(4 \gamma_m )\right)\versorz$ obtained by applying $2\Ptrot$ successive rotations to $\versorz$. 
Therefore $\epsilon_{k}$ assumes its minimum value $0$ when 
\begin{equation} \label{eqn:constraints}
\left(\Tprod{\Ptrot}_{m=1}\Rrot_{\versorz}(4 \beta_m )\Rrot_{\versorb_k}(4 \gamma_m )\right) \versorz = \versorb_k \;.
\end{equation} 
Eq.~\ref{eqn:constraints} represents a set of {\em constrains} for each wave-vector $k$. If all constrains can be satisfied simultaneously,  the minimal residual value of the residual energy can be obtained by setting $\epsilon_{k}=0$ in Eq.~\ref{eqn:eres_abc} and Eq.~\ref{eqn:eres_pbc}. We find that by 
the number of free variational parameters and the number of constraint equations one can get a picture of the QAOA landscape. 
Figure~\ref{fig:QAOA_minima} illustrates the role of these counting arguments for $\Ptrot=3$.
Here we observe that:
\begin{description}
    \item[For $\mathbf{2P<N}$:] the number of constraints $2|\calK_{\ABC}|=2\Ptrot$ --- corresponding to $\Ptrot$ equations (\ref{eqn:constraints})
    for $3$-dimensional unit vectors --- and of variables $2\Ptrot$
    are equal. Therefore the equations have a finite set of discrete solutions. 
    When all equations are satisfied we get, see \eqref{eqn:eres_abc}, the optimal $\eres(\bgamma^*,\bbeta^*)=(2\Ptrot+2)^{-1}$.
    \item[For $\mathbf{2P=N}$:] again the number of constraints $2|\calK_{\PBC}|=N$ and of variables $2\Ptrot$ are equal. 
    Therefore the equations still have discrete solutions. When all equations are satisfied, the residual energy, see \eqref{eqn:eres_pbc}, 
    is $\eres(\bgamma^*,\bbeta^*)=0$. 
    \item[For $\mathbf{2P>N}$:] the number of constraints $2|\calK_{\PBC}|=N$ is smaller than the number of variables $2\Ptrot$. 
    The equations therefore have a continuous set of solutions that define a manifold of dimension $2\Ptrot - N$. 
\end{description}

\begin{figure}
    \includegraphics[width=\columnwidth]{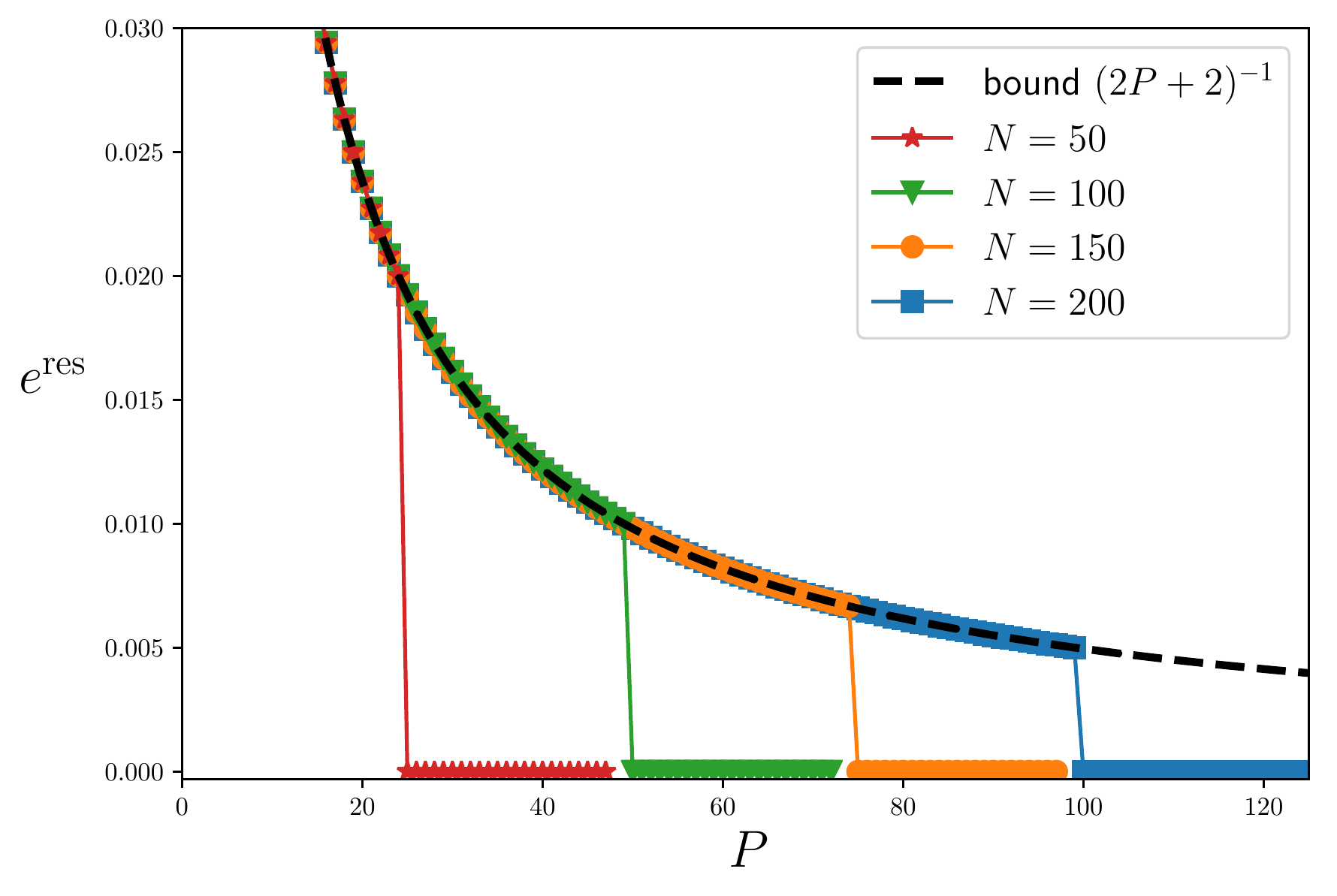}
\caption{Optimal residual energies $\eres$ obtained applying the QAOA with $\Ptrot$ Trotter steps  for various system sizes $N=50, 100, 150, 200$.  
The symbols represent the data obtained by numerical optimization while the dashed black line represents the theoretical bound obtained in Eq.~\eqref{eqn:eres_final}, 
which holds for $2\Ptrot<N$. The bound is saturated for $2\Ptrot<N$ while for $2\Ptrot\geq N$ the residual energy drops to zero due to existence of the 
solution described by Eq.~\eqref{eqn:opt_sol_P=N/2}.}
    \label{fig:eres_bound_vs_P}
\end{figure}
Figure~\ref{fig:eres_bound_vs_P} illustrates the minimum residual energy obtained for different values of 
$N$ as a function of $\Ptrot$. 
The numerical data are obtained by looking for optimal solutions via a numerical minimization of the 
residual energy, Eq.~\eqref{eqn:eres_abc} and Eq.~\eqref{eqn:eres_pbc}, using Eq.~\eqref{eqn:eresk_geometrical_def} to compute the terms $\epsilon_k$. 
%
%
%
Specifically we implement the function $\eres_{\Ptrot}(\bgamma,\bbeta)$ with \texttt{PyTorch} that provides built-in auto-differentiation 
routines~\cite{Paszke_NIPS2017}. 
We then minimize $\eres_{\Ptrot}(\bgamma,\bbeta)$ with the Broyden-Fletcher-Goldfard-Shanno (BFGS) optimization algorithm~\cite{Nocedal_book2006}, 
using back-propagation to compute the required gradients. The algorithm is halted when the residual energy is sufficiently close 
--- specifically, within $10^{-7}$ --- to our theoretical lower bound $\eres_{\Ptrot}=\frac{1}{2\Ptrot+2}$. 


The global minima returned by the BFGS routine depend on the arbitrary choice of the initial guess for $\bgamma$ and $\bbeta$. 
In particular, we conjecture (and have verified numerically for $P\leq6$) that there are $2^{\Ptrot}$ {\em degenerate minima} all sharing the same 
$\eres_{\Ptrot} = \frac{1}{2\Ptrot+2}$ for $N> 2\Ptrot$. 
Notice that $\eres_{\Ptrot}$ drops to $0$ when $2\Ptrot\ge N$, as predicted by the parameter counting argument presented above. 

In Ref.~\onlinecite{Mbeng_optimal-dQA_arXiv2019} we have demonstrated that, among all these degenerate solutions, one can single-out a rather special {\em regular solution} 
which is closely related to the problem of an optimal QA \cite{Zanardi_PRL2009}.

\subsection{Optimal schedules for $2\Ptrot\! < \!N$ and digitized-QA} \label{sec:from_QAOA_to_QA}
As discussed previously, for $2\Ptrot<N$ the QAOA landscape is independent on the system size, and one is effectively considering an
infinite chain $N\to\infty$. 
There  are various equivalent optimal choices for the $\gamma_m$ and $\beta_m$, most of which lack any structure or pattern. 
Here, we illustrate the construction \cite{Mbeng_optimal-dQA_arXiv2019} of a \emph{regular} schedule that shows a well defined continuous limit when $\Ptrot \to \infty$.

\begin{figure}
    \includegraphics[width=\columnwidth]{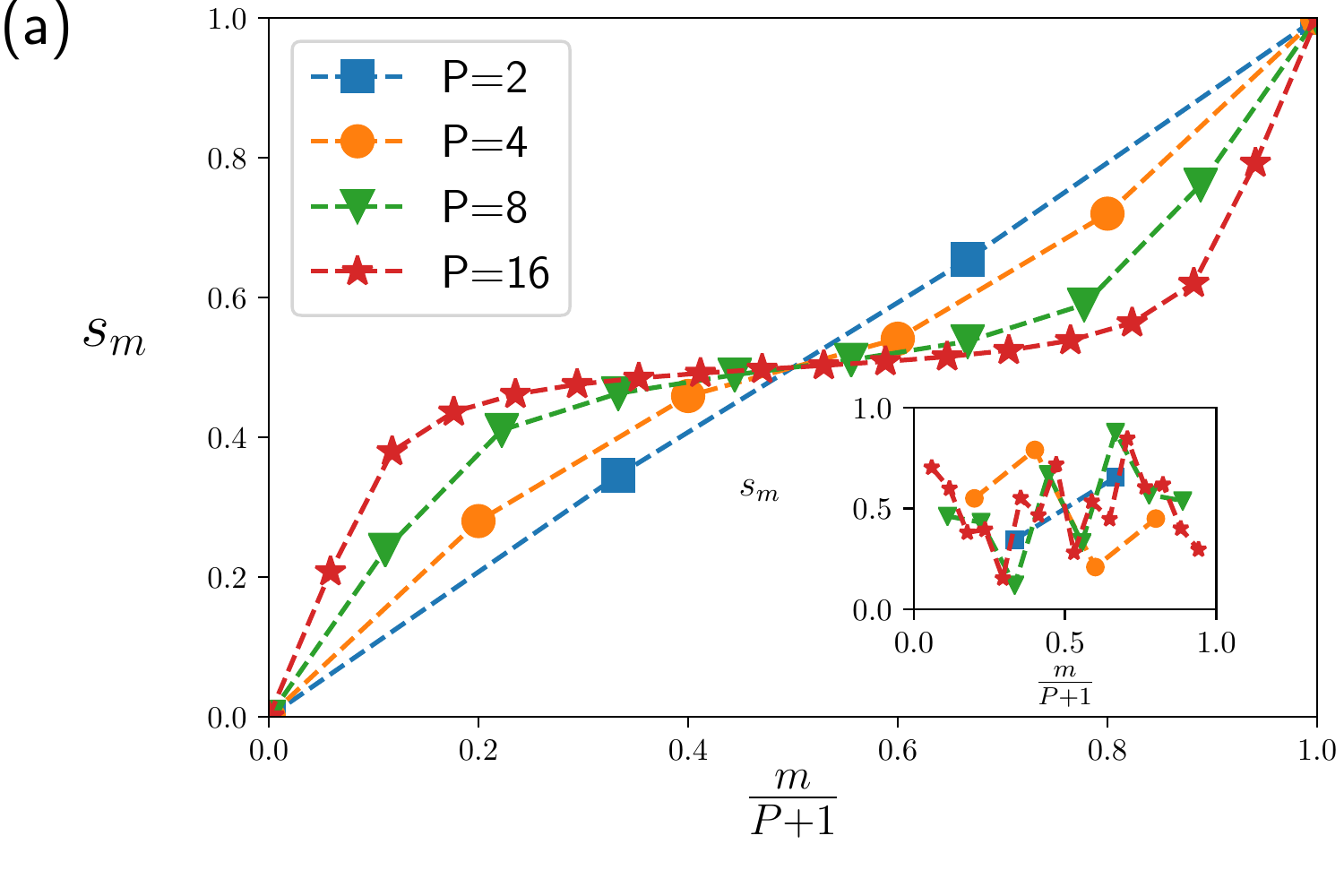}
    \includegraphics[width=\columnwidth]{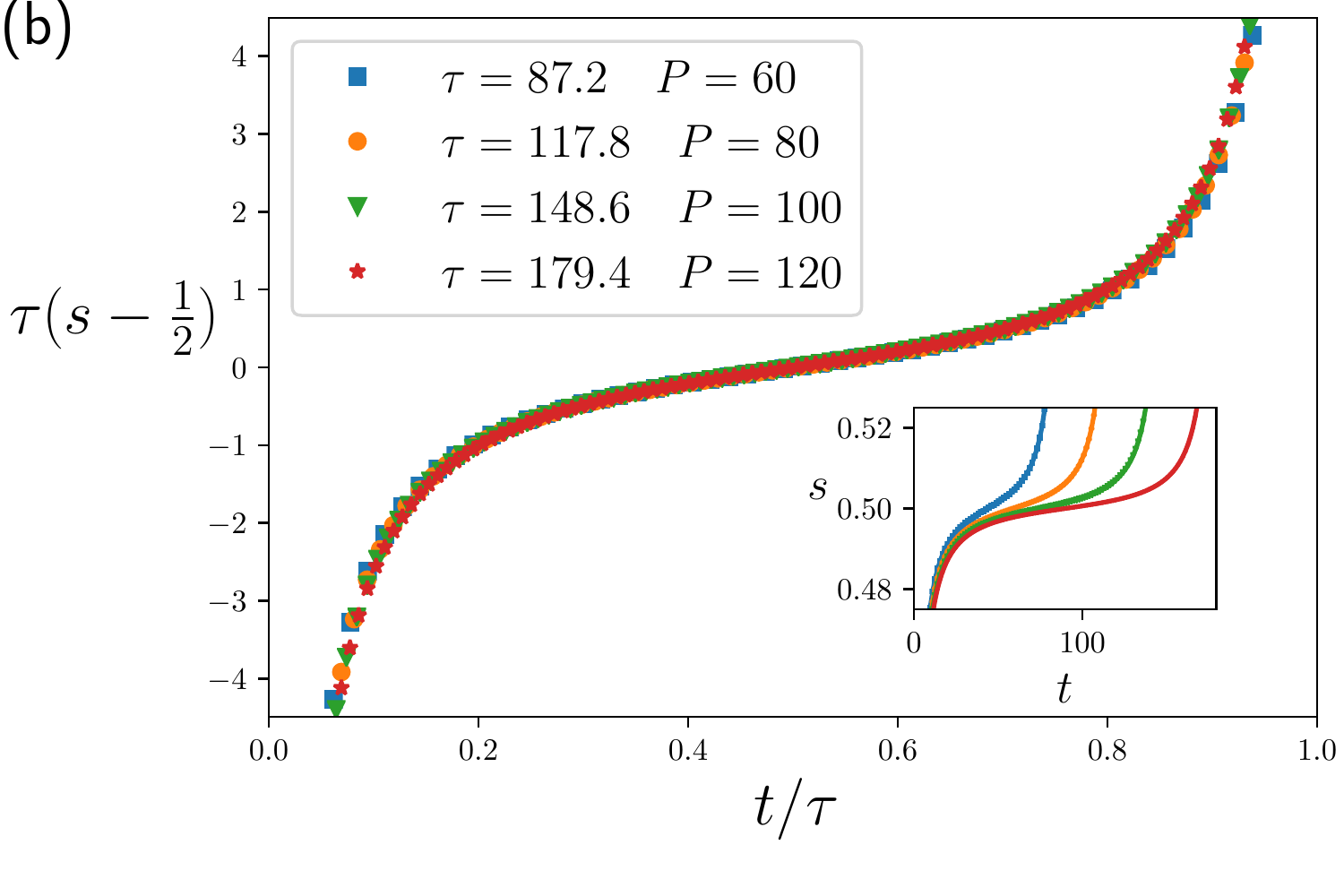}
    \caption{ (a) Construction of the ``regular'' solution for increasing $\Ptrot$. 
        We plot here the parameter 
        $s_m = \frac{\gamma_m}{\gamma_m + \beta_m}$. 
        The regular solution for $\Ptrot$ Trotter steps is obtained using as initial guess for the local search algorithm the solution obtained for a smaller value of $\Ptrot$  
        (e.g. $\Ptrot/2$). This procedure produces a smooth schedule. 
        The inset shows a set of generic optimal solution obtained when initializing the QAOA angles randomly. 
        This procedure apparently produces irregular patternless solutions. 
        (b) Control protocol $s(t)$ induced by the QAOA algorithm for several values of the total time 
        $\Tann$ (or equivalently the number of Trotter steps $\Ptrot$). The protocols have been scaled according 
        to Eq.~\eqref{eq:scaling_law}. The inset shows the unscaled protocols.}
    \label{fig:regular_schedule}
\end{figure}

To specifically target the regular solution, we proceed iteratively in $\Ptrot$. 
As proposed in Ref.~\onlinecite{Mbeng_optimal-dQA_arXiv2019}, we obtain the optimal solution at level $\Ptrot$ by using as initial guess for $\bgamma, \bbeta$ 
the regular solution previously obtained at level $\Ptrot'<\Ptrot$. The construction is illustrated in Fig.~\ref{fig:regular_schedule}.
The starting point is a linear digitized QA schedule $(\bgamma^{\mathrm{lin}},\bbeta^{\mathrm{lin}})_{\Ptrot=2}$, which is obtained by a $\Ptrot=2$ digitalization of the QA schedule 
$s(t)=t/\tau$. We  use $(\bgamma^{\mathrm{lin}},\bbeta^{\mathrm{lin}})_{\Ptrot=2}$ as initial guess for  a BFGS search and find the optimal parameters 
$(\bgamma^{\regular},\bbeta^{\regular})_{\Ptrot=2}$. 
To visualize the optimal protocols, we invert Eq.~\eqref{eqn:gamma_beta_m} and plot the effective parameter 
\begin{equation}
s_m=\frac{\gamma_m}{\gamma_m+\beta_m}\;.
\end{equation}
Fig.~\ref{fig:regular_schedule} shows that, for $\Ptrot=2$ the optimal protocol
is still close to the original linear digitized QA.
We next consider $\Ptrot=4$ and use $(\bgamma^{\regular},\bbeta^{\regular})_{\Ptrot=2}$ as initial guess for a new BFGS search. 
In Fig.~\ref{fig:regular_schedule} we visualize the new optimal protocol $(\bgamma^{\regular},\bbeta^{\regular})_{\Ptrot=4}$, which now deviates from the linear digitized QA. 
Iterating this procedure leads to the {\em smooth optimal protocols} shown in Fig.~\ref{fig:regular_schedule}. 
On the other hand, the inset of Fig.~\ref{fig:regular_schedule} illustrates the ``irregular" values of $s_m$ resulting from a random initialization of the BFGS search. 
Such a patternless and irregular behavior is common to the vast majority of optimal solutions. However,  the iterative search scheme outlined, enables us to single out a {\em regular} protocol that exhibit a well recognizable ``structure''. In Fig.~\ref{fig:regular_schedule}(b) we analyze this structure further, by studying the schedule parameters $s_m$.
The inset of Fig.~\ref{fig:regular_schedule}(b) shows that by increasing $\Ptrot$, 
hence the total $\Tann$ given by Eq.~\eqref{eqn:sum_rule}, the schedule $s_m$, when expressed in terms of the corresponding time
$t_m = \sum_{m'=1}^m (\gamma_{m'}+\beta_{m'})/\hbar$, appears to become flatter and flatter close to the quantum critical point $s=1/2$. 
Remarkably, the whole regular solution shows a simple scaling of the form 
\begin{equation} \label{eq:scaling_law}
s_{\Tann}(t) = \frac{1}{2} + \frac{1}{\Tann^{\alpha}} f\left( \frac{t}{\Tann} \right)
\end{equation}
with $\alpha=1$, as illustrated by the data collapse in the main plot of Fig.~\ref{fig:regular_schedule}(b).
Such a shape of $s(t)$ is clearly reminiscent of the adiabatic protocols described in the context of a 
continuous-time QA in Ref.~\onlinecite{Roland_PRA2002} or Ref.~\onlinecite{Polkovnikov_PRL2008}. 
We will further comment on such an issue in Sec.~\ref{subsec:comparison}. 

We now explore the connection with QA, or more properly to a form of {\em digital-QA}~\cite{Martinis_Nat16}.
The optimal parameters of the regular solution $(\bgamma^{\regular},\bbeta^{\regular})$ define a candidate digitized-QA schedule, from which,
by inverting Eq.~\eqref{eqn:gamma_beta_m},  one can construct an associated step-QA and a continuous-time QA protocol, as illustrated in 
Fig.~\ref{fig:bang-bang} for $\Ptrot=8$.




The iterative construction targets an optimal solution $\bgamma^\regular,\bbeta^\regular$ that varies weakly from $\Ptrot$ to $\Ptrot+1$. 
However, to explore the connection with a digitized-QA, we also need to verify and quantify the {\em adiabaticity} of the dynamics defined by $\bgamma^\regular,\bbeta^\regular$.
 
During the preparation of the variational state given in Eq.~\eqref{eqn:psi_qaoa}, the system undergoes a unitary discrete time evolution. 
The intermediate state $\ket{\psi_{m}( \bgamma,\bbeta) }$ after $m$ steps satisfies the following discrete version of Schr\"odinger's equation
\begin{eqnarray}
\ket{\psi_{0}(\bgamma,\bbeta) } &=& \ket{\psi_0}\\
\ket{\psi_{m+1}( \bgamma,\bbeta) }&=& \Uevol_{m} \ket{\psi_{m}(\bgamma,\bbeta) }
\end{eqnarray}
where we recall that the effective discrete time evolution operator is defined to be
$\Uevol_{m}=\Uevol(\gamma_m,\beta_m)=\ee^{-i \beta_{m} \Ham_{x} }\ee^{-i\gamma_{m} \Ham_z}$.
We can always find an orthonormal basis $\ket{\theta_{m}}$ that diagonalizes $\Uevol_m$:
\begin{equation}
\Uevol_m\ket{\theta_{m}} =\ee^{-i\theta_{m}}\ket{\theta_{m}}
\end{equation}
%
%
We say that $(\bgamma,\bbeta)$ defines an adiabatic dynamics if the state $\ket{\psi_{m}( \bgamma,\bbeta) }$ closely {\em follows an eigenstate} 
$\ket{\overline{\theta}_{m}}$ of $\Uevol_m$. 
This is a natural extension \cite{Dranov_JMatPhys1998} of the concept of adiabadicity in continuous-time dynamics. 
For instance, one can show that a digitized-QA schedule obtained by discretizing  a continuous-QA 
({\em e.g.}, using Eq.~\eqref{eqn:Trotter_lowest} and Eq.~\eqref{eqn:gamma_beta_m}) is as adiabatic as its continuous counterpart when $\Ptrot\to\infty$. 
Moreover this definition is further justified by the statements in Ref.~\onlinecite{Dranov_JMatPhys1998}, where a discrete version of the adiabatic theorem is given.
 
\begin{figure}
    \includegraphics[width=\columnwidth]{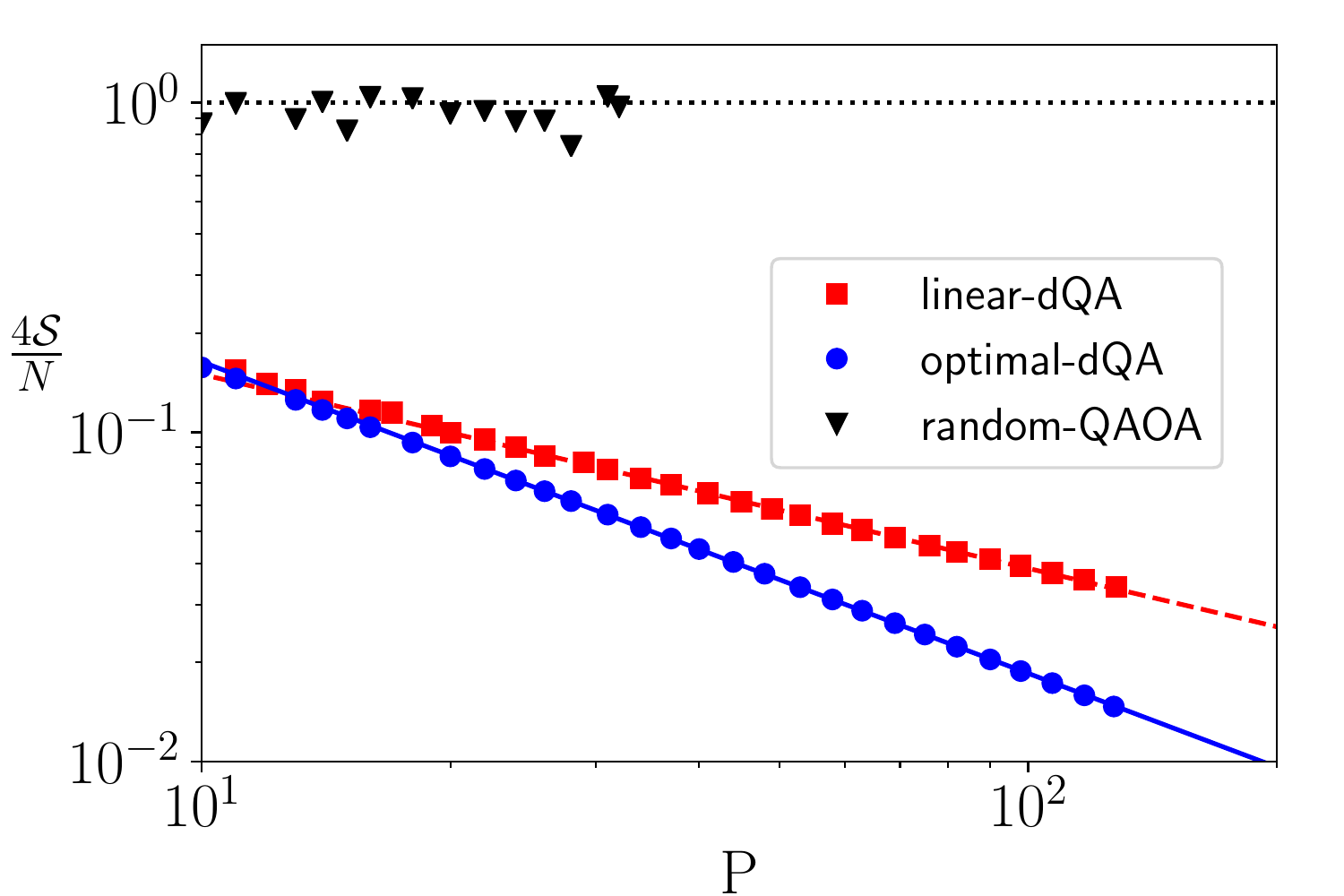}     
    \caption{Average Shannon entropy $\calS_{ \bgamma,\bbeta}$ defined in Eq.~\eqref{eqn:Q_shannon}, normalized to $N/4$, for various schedules.}
     \label{fig:QS_adiabaticity}
\end{figure}

Let  $p_{ \bgamma,\bbeta}(\theta_{m})=|\braket{\theta_{m}}{\psi_{m}( \bgamma,\bbeta)}|^2$ be the probability of finding the system in a given 
eigenstate $\ket{\theta_{m}}$. 
The definition of adiabatic dynamics given above, suggests to quantify the degree of adiabaticy by measuring how close the distribution 
$p_{ \bgamma,\bbeta}(\theta_{m})$ is to a degenerate one (i.e., a Kr\"oneker-delta in $\theta_{m}$). 
The adiabaticity of the discrete dynamics with $P$ steps, can then be quantified with the average Shannon entropy 
$\calS_{ \bgamma,\bbeta}(\Ptrot) $ of the distribution $p_{ \bgamma,\bbeta}(\theta_{m})$:
\begin{equation}\label{eqn:Q_shannon}
\calS_{ \bgamma,\bbeta}(\Ptrot) =-\frac{1}{\Ptrot} 
\sum_{m=1}^\Ptrot\sum_{\theta_{m}}p_{ \bgamma,\bbeta}(\theta_{m})\log [p_{ \bgamma,\bbeta}(\theta_{m})]\,
\end{equation}
%
For an adiabatic dynamics $\calS_{ \bgamma,\bbeta}(\Ptrot)\to 0$ as $\Ptrot\to\infty$, otherwise it should remain finite. 
In Fig.~\ref{fig:QS_adiabaticity} we show such Shannon entropy for three different schedules on an Ising chain with $N=1024$ sites. 
We first take a look at $\calS_{ \bgamma,\bbeta}(\Ptrot)$ for a linear digitized-QA schedule ($\Delta t_m = 1$) which is represented by red squares. 
Through the adiabatic theorem we know that the linear-digitized schedule is adiabatic for $\Ptrot\to\infty$ and $\calS_{ \bgamma,\bbeta}(\Ptrot)$, 
decaying to zero when $\Ptrot$ is increased, correctly signals the emerging adiabaticity of the schedule. 
We then consider a generic optimal solution found by the QAOA algorithm staring form a random initialization  (black triangles). 
We find that $4\calS_{ \bgamma,\bbeta}(\Ptrot)/N\approx1$ independently of $\Ptrot$, signalling a non-adiabatic dynamics. 
Finally, the blue circles were obtained from the regular QAOA solution $\bgamma^\regular,\bbeta^\regular$. 
The fact that $\calS_{ \bgamma^\regular,\bbeta^\regular}(\Ptrot)$ vanishes as $\Ptrot\to\infty$ immediately conveys the message that the regular QAOA 
solution defines an adiabatic schedule. 
Moreover $\calS_{ \bgamma,\bbeta}(\Ptrot)$ allows us to make quantitative statements: 
In particular, the regular QAOA solution is evidently {\em more} adiabatic than the linear digitized-QA schedule. 
We conclude that such optimal solution can be interpreted as an improved adiabatic digitized-QA schedule. 
In the App.~\ref{app:Heff}, we discuss how a suitable effective Hamiltonian can be introduced for the digitized-QA. 

\subsection{Optimal schedules for $2\Ptrot\! \ge \! N$ and Quantum Control}
\begin{figure}
     \includegraphics[width=\columnwidth]{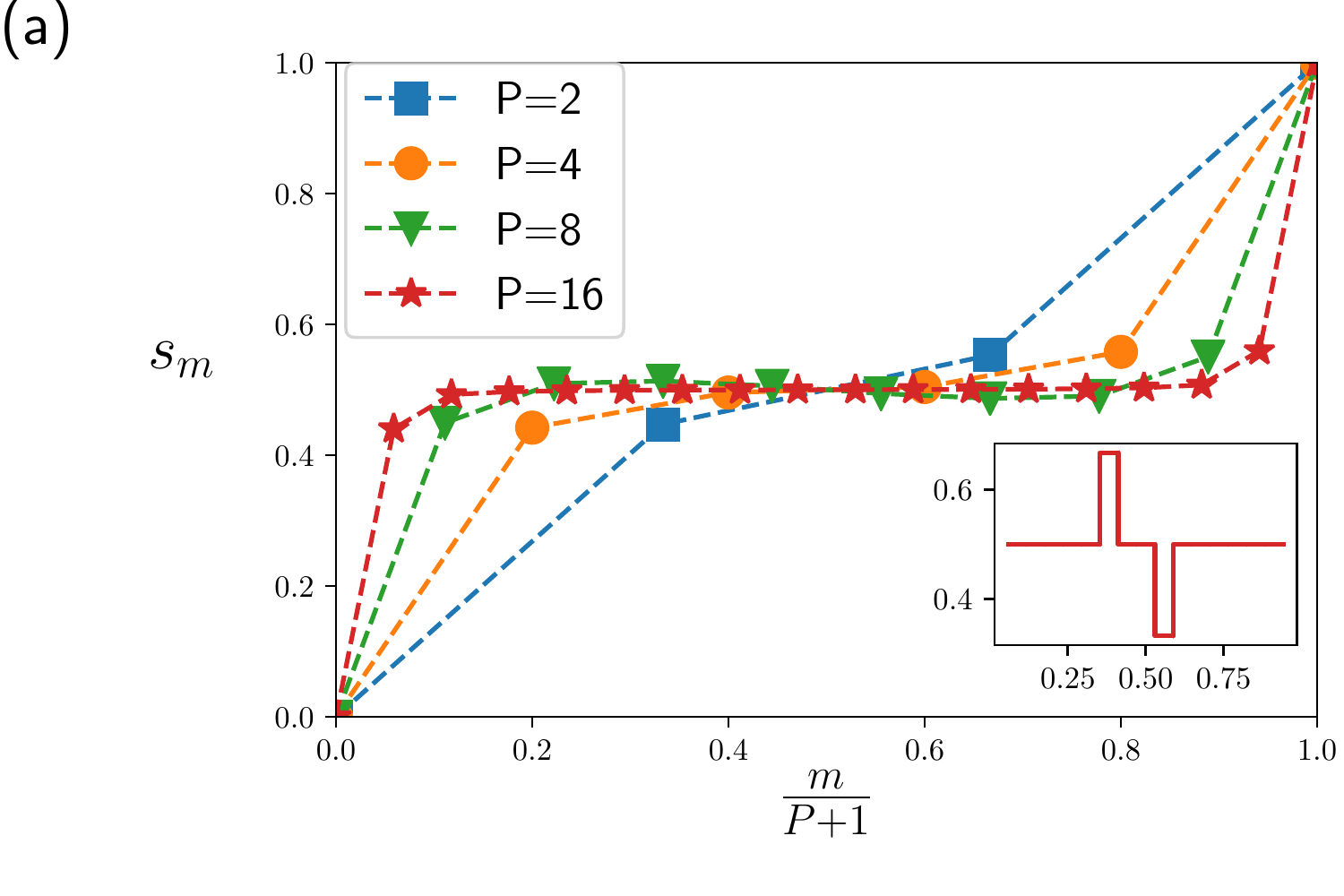}
     \includegraphics[width=\columnwidth]{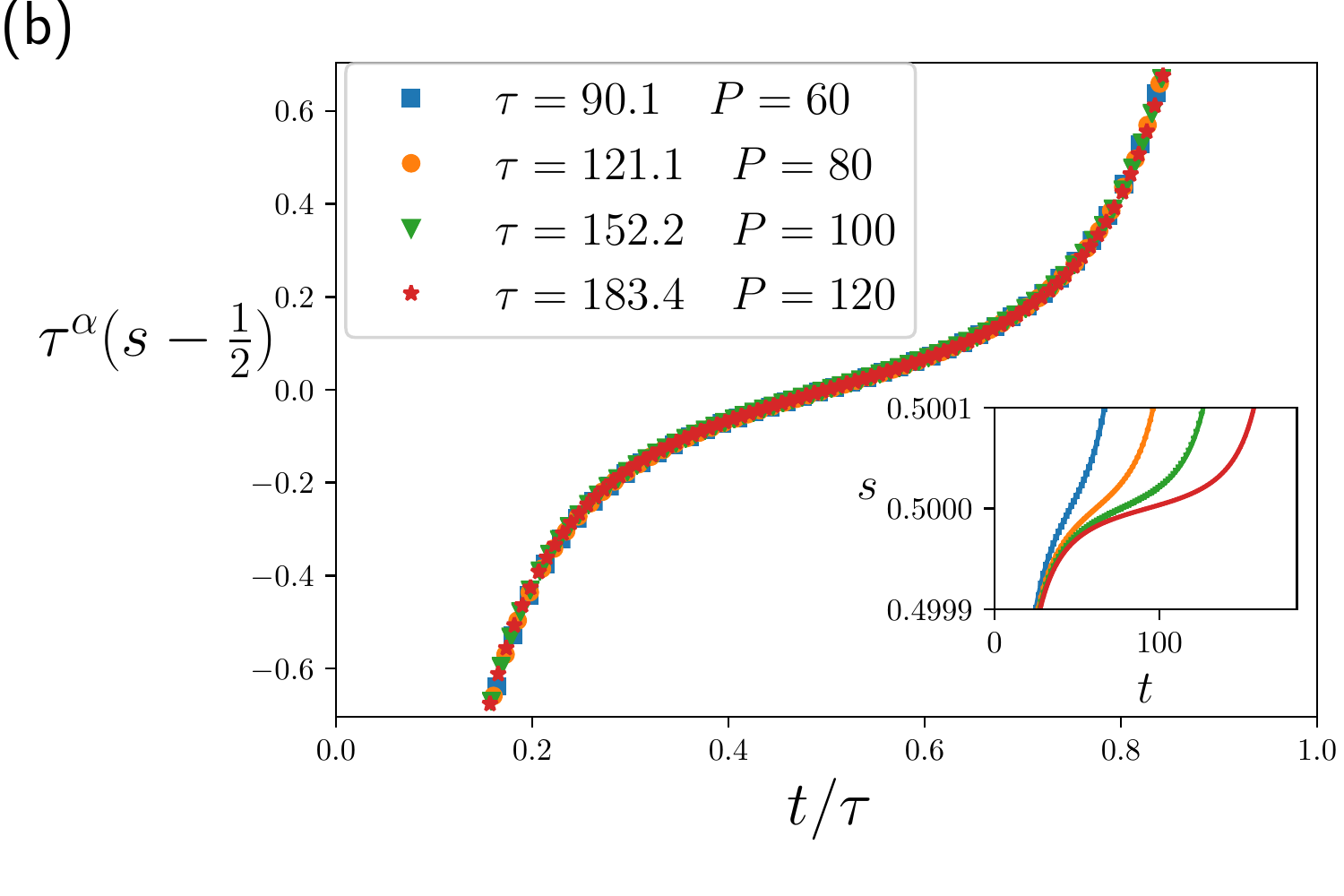}
    \caption{(a) Construction of the ``regular'' solution for increasing $\Ptrot$ when $2\Ptrot\ge N$. 
    We plot here the parameter $s_m = \frac{\gamma_m}{\gamma_m + \beta_m}$. 
    The regular solution for $\Ptrot$ Trotter steps is obtained using as initial guess for the local search algorithm the solution obtained for a smaller value of $\Ptrot$  
    (e.g. $\Ptrot/2$). This procedure produces a smooth schedule. 
    The inset shows the optimal solution in Eq.~\eqref{eqn:opt_sol_P=N/2}.
    (b) Control protocol $s(t)$ induced by the QAOA algorithm for several values of the total time 
        $\Tann$ (or, equivalently, number of Trotter steps $\Ptrot$). The protocols have been scaled according 
        to Eq.~\eqref{eq:scaling_law} with $\alpha=1.75$. The inset shows the unscaled protocols.}
    \label{fig:s_reg_eres0}
\end{figure}

For $2\Ptrot\geq N$ it is always possible to prepare the Ising $\Ham_z$ ground state with a QAOA {\em Ansatz}: the system is controllable. 
This can be done by explicitly showing that the specific schedule
\begin{equation}\label{eqn:opt_sol_P=N/2}
\beta_m =\gamma_{\Ptrot+1-m}=
\begin{cases}
\pi / 8 \,\,\mbox{ if } m = \ceil{\frac{\Ptrot+1}{2}}\vspace{2mm} \\
\pi / 4\,\, \mbox{ otherwise} 
\end{cases}
\end{equation}
realizes exactly $\eres_{\Ptrot}=0$. 
The rationale behind such a remarkably simple expression is that most of the rotations  $\Rrot_{\versorz}(4 \beta_m )\Rrot_{\bvec_k}(4 \gamma_m )$
involved in Eq.~\eqref{eqn:eresk_geometrical_def} are rotations by $\pi$ and their combined effect leave the vector $\versorb_k$ in the same plane as $\versorz$  
while shifting the angle between them by $2k$. The discrete nature of the $k$-vectors involved guarantees that one effectively rotates,
by using the angles in Eq.~\eqref{eqn:opt_sol_P=N/2}, each $\versorb_k$ onto $\versorz$, as a detailed construction (not given here) shows.  
More generally, however, since the problem is now underdetermined (the number of variational parameters is larger than the
number of constraints), one can construct a {\em continuum} of optimal solutions. 
In particular, using the same iterative strategy described in the previous section, we can single-out a regular solution attaining 
$\eres_{\Ptrot}(\bgamma^\regular,\bbeta^\regular)=0$. 
Figure~\ref{fig:s_reg_eres0}(a) shows the construction of such a regular schedule for $2\Ptrot=N$.  
%
%
%
Concerning a collapse of the data, we verified that the {\em Ansatz} in Eq.~\eqref{eq:scaling_law} still works, but now with a modified exponent
$\alpha\approx 1.75$. 
On the practical side, observe that the schedule becomes flatter and flatter across the quantum critical point ($s=1/2$). 

 
\subsection{Comparison with other QA strategies} \label{subsec:comparison}
We have discussed how the QAOA framework can be used to construct a regular schedule that realizes an optimal digitized-QA dynamics, 
without relying on any knowledge of spectral information.
In this section we compare the performance of such optimal regular digitized-QA schedule with the performance of other standard QA schedules for the translationally 
invariant Ising chain problem. 
Specifically, we compare the scaling of the residual energy $\eres(\tau)$~\cite{Santoro_SCI02} for various continuous time-schedules and their digitized counterparts.

\begin{figure}
    \includegraphics[width=\columnwidth]{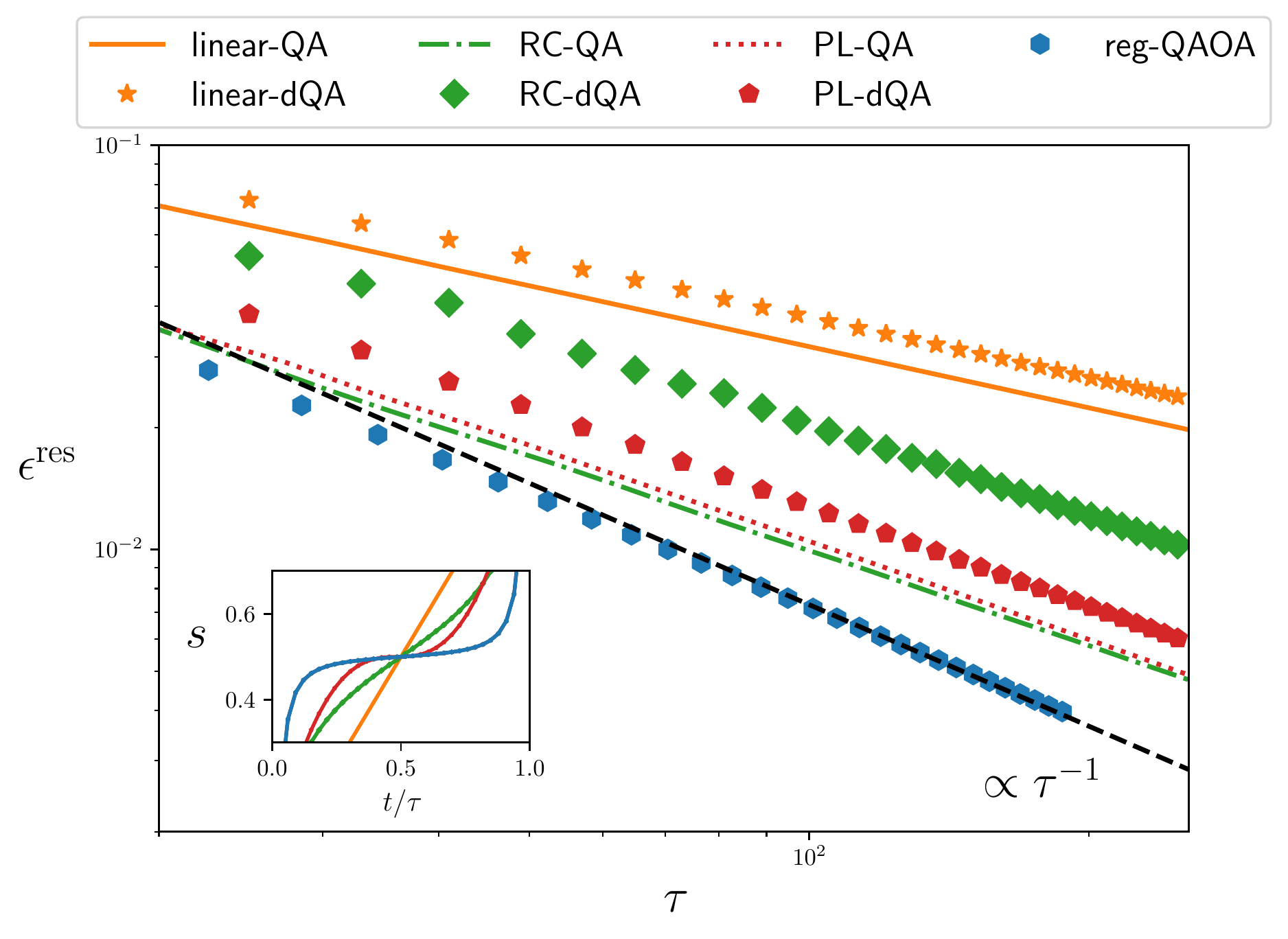}
    \caption{Scaling of the optimal defects for various QA schedules applied to the infinite Ising chain problem. 
    All digitized-QA data assume a Trotter discretization with $\Delta t_m=1$. 
    The linear continuous-time QA (orange solid solid line) and digitized-QA (orange stars) show a Kibble-Zurek exponent $\eres\sim \Tann^{-1/2}$. 
    The Roland-Cerf QA and dQA (green dash-dot line  and diamonds) and the power-law QA and dQA (red dotted line and pentagons) with optimized parameters
    show  $\eres\sim \Tann^{-\alpha}$ with $\alpha\approx 0.75$ and $\alpha\approx 0.8$, respectively. 
    The brown triangles represent the optimal QAOA regular results. 
    The inset shows the values of $s(t)$ for fixed $\tau=32$ for the different schedules.}
    \label{fig:QAOA_ndef}
\end{figure}

The standard and most natural choice in analog QA, is a linear-schedule continuous-time QA (linear-QA), $s(t)=t/\Tann$. 
In the translationally invariant Ising chain problem~\cite{Dziarmaga_PRL05,Zurek_PRL05}, the scaling of the residual energy for a linear-QA schedule is ruled by the Kibble-Zurek 
(KZ)~\cite{kibble76, zurek85, Polkovnikov_RMP11} mechanism, which predicts a power-law scaling $\eres(\Tann) \sim \Tann^{-1/2}$. 
We can digitize the linear-QA schedule, for instance with $\Delta t_m=1$ (in units of $\hbar/J$)~\cite{Mbeng_dQA_PRB2019}.  
The result of this operation is a linear digitized-QA (linear-dQA) schedule, which is suitable to run on a digital quantum hardware. 
Figure \ref{fig:QAOA_ndef} shows that the same KZ scaling $\eres(\Tann) \sim \Tann^{-1/2}$ holds for both linear-QA and linear-dQA~\cite{Mbeng_dQA_PRB2019}. 
The only visible effect of digitalization is to separate the two curves by a constant off-set.

Next, we consider other optimized schedules that have been proposed in the context of continuous-time QA. 
One was proposed in \cite{Roland_PRA2002}, where $s(t)$ has the form:
\begin{equation}
s(t) = \frac{1}{2} + \frac{1}{2C} \tan \left( 2 \left(\frac{t}{\Tann}-\frac{1}{2}\right) \arctan(C)  \right) 
\end{equation}
where $C$ is a parameter determining the slope at the critical point $s_c=1/2$. 
In general, $C$ depends on $\tau$, and should be optimized.
%
%
Alternatively, one can consider a power-law schedule, as proposed in \cite{Polkovnikov_PRL2008}:
\begin{equation}
s(t) = \frac{1}{2} + \frac{1}{2} \mathrm{sgn} \left( \frac{t}{\Tann}-\frac{1}{2}\right)  \left| 2 \frac{t}{\Tann}-1\right|^C 
\end{equation}
$C$ being now the power-law exponent,
again dependent in general on $\tau$ and to be optimized.
%
%
Both these strategies exploit the knowledge of the critical point location, here at $s_c=1/2$, and can be applied either within a continuous-time QA, 
or, after digitalization, as dQA.
Numerically, they both produce an improvement over linear-QA, with $\eres \sim \Tann^{-\alpha}$, where 
$\alpha\sim 0.75$ and $\alpha\sim 0.8$.
In all cases, the digitalization appears to add a constant offset {\em upwards} to the continuous-time curves, with identical power-law exponent.
This seems to be at variance with what the Trotter error does in Simulated Path-Integral Monte Carlo QA~\cite{Santoro_SCI02, Heim_SCI15, Mbeng_PRB2019}.   

Finally, Fig.~\ref{fig:QAOA_ndef} shows the residual energy corresponding to the optimal digitized-QA solution, with $\tau$ calculated from \eqref{eqn:sum_rule}. 
Here the behaviour of $\eres(\Tann)$ shows the optimal power-law $\eres\sim \Tann^{-1}$, coherently with the bound $\eres_{\Ptrot}\ge (2\Ptrot+2)^{-1}$
and with $\tau \propto \Ptrot$. 

\begin{figure}
    \includegraphics[width=\columnwidth]{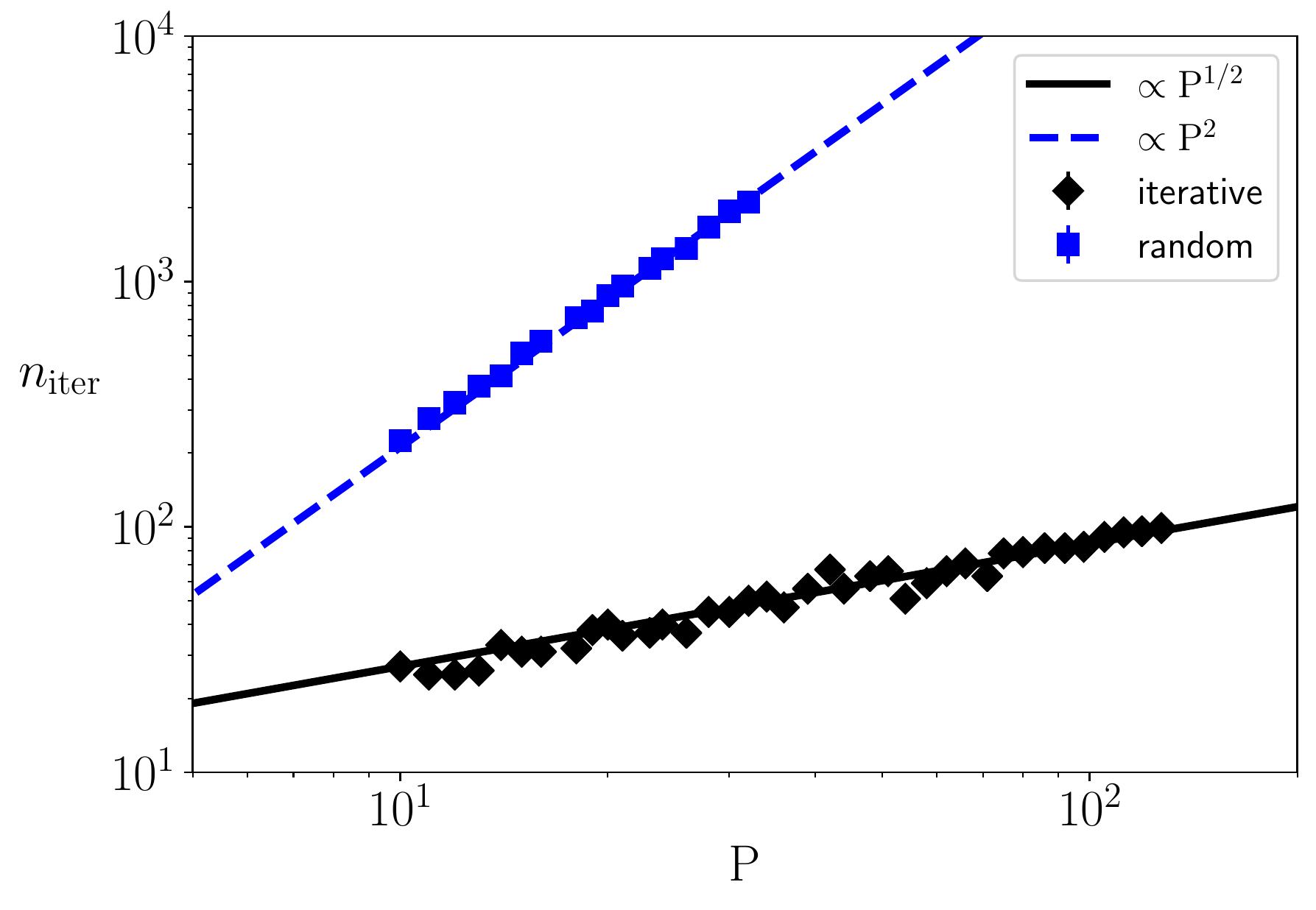}
    \caption{
        Number of iterations needed by the QOAO algorithm to converge to a minimum, with tolerance $10^{-5}$. 
        The black diamonds refer to the iterative search where the search system is initialized by interpolating a solution 
        obtained for a lower value of $\Ptrot$, while the blue squares represent a brute-force search starting from a random initial point. 
        The classical optimization is performed using the BFGS algorithm~\cite{Nocedal_book2006}.
    }
    \label{fig:QAOA_cpu_cost}
\end{figure}

The regular optimal dQA solution has the best possible performance, saturating the residual energy bound: $\eres\sim \Tann^{-1}$. 
However, such a quadratic speed-up over the plain KZ exponent comes with an extra computational cost to find the global QAOA variational minimum. 
Figure~\ref{fig:QAOA_cpu_cost} shows that the number of iterations $n_{\rm iter}$ to find a minimum by starting from a random initial point increases as $\Ptrot^2$,
while $n_{\rm iter}\propto \sqrt{\Ptrot}$ for the iterative search of the regular optimal dQA solution.  

Let us estimate how the residual energy decreases as a function of the computational cost $\tcomp$. 
One of the issues is the computational cost associated with a call of the ``quantum oracle''. 
Suppose we agree that such a cost scales with $\Ptrot$, the number of unitaries involved in preparing $\ket{\psi_\Ptrot(\bgamma,\bbeta)}$, 
so that $\tcomp\propto \Ptrot n_{\rm iter}$.  
Then, the linear-dQA has $\tcomp\propto \Ptrot$, the random search of the optimal solution has
$\tcomp \propto n_{\rm iter} \, \Ptrot \propto \Ptrot^3$, and the iterative search of the optimal dQA solution has 
$\tcomp \propto n_{\rm iter} \, \Ptrot \propto \Ptrot^{3/2}$. 
Using these estimates, we can express the residual energies in terms of the computational cost: 
\begin{equation}
\left\{ \begin{array}{ll}
\eres \sim \Ptrot^{-\frac{1}{2}} \sim \tcomp^{-\frac{1}{2}} & \hspace{2mm} \mbox{(linear-dQA)} \vspace{2mm}\\
\eres \sim \Ptrot^{-1} \sim \tcomp^{-\frac{1}{3}} & \hspace{2mm} \mbox{(QAOA, random)} \vspace{2mm} \\
\eres \sim \Ptrot^{-1} \sim \tcomp^{-\frac{2}{3}} & \hspace{2mm} \mbox{(optimal-dQA, recursive)} 
\end{array}
\right. \;.
\end{equation}
Hence the overall performance of the optimal QAOA for a random initialization, in terms of computational time, is definitely {\em worse} than plain linear-dQA.  
To improve over linear-dQA, one {\em must} use a recursive initialization, leading to an optimal-dQA. 
%

\section{Discussion and conclusions} \label{sec:conclusions}
We have discussed the links between Quantum Annealing (QA), both in its continuous-time version and in its
digital flavour, with the hybrid quantum-classical variational approach known as QAOA, elucidating the connection between
optimal Quantum Control and the requirement of adiabaticity of the driving protocols.

Two are the main contributions we have discussed. 
The first is a technique to establish a variational bound on the residual energy of MaxCut problems on $2$-regular periodic graphs
by playing with the boundary conditions on the reduced spin problem. 
Such a technique can be naturally extended to higher-dimensional problems~\cite{Mbeng_PhDThesis2019}, and allows, through the use of Lieb-Robinson bounds, 
the physical evolution time to enter the game. We will deal with these issues in a separate publication,
discussing also the role of entanglement and the presence of a light-cone, associated with a local Hamiltonian. 
In one dimension, we have shown that the variational bound $\eres_{\Ptrot}\ge (2\Ptrot+2)^{-1}$ is precisely saturated by the Jordan-Wigner results, 
which also helps to elucidate the geometric nature of the minimization problem and the role of the variational parameters, $2\Ptrot$, in 
comparison with the number of spins, $N$. 
This, in turn, shows that the system becomes controllable, and the residual energy drops to $0$, as soon as $2\Ptrot\ge N$.

The second contribution~\cite{Mbeng_optimal-dQA_arXiv2019} has to do with the link between Quantum Control, which generally predicts the optimal schedule to be 
of the bang-bang form~\cite{Yang_PRX2017}, hence justifying the QAOA {\em Ansatz}~\cite{Farhi_arXiv2014}, and the adiabatic dynamics 
behind QA, or more precisely here {\em digitized}-QA~\cite{Martinis_Nat16}.
Indeed, among a large number of QAOA optimal solutions --- $2^\Ptrot$ for $2\Ptrot < N$, a continuum for $2\Ptrot>N$ ---
one can iteratively single-out a {\em smooth regular} solution which can be regarded as the optimal digitized-QA schedule. 
%
Such a regular optimal solution provides a clear speed-up over linear-QA. 
The speed-up is quadratic --- as in the Grover problem \cite{Roland_PRA2002} --- if the computational cost for finding the solution is not considered. 
The speed-up still survives even when we account for the cost of searching the minimum, but only if smart iterative techniques~\cite{Mbeng_optimal-dQA_arXiv2019}
to construct the optimal solutions are used.

One point which is worth remarking is that the smooth-regular-adiabatic digitized-QA solution that we construct {\em does not use} any prior knowledge 
on the location of the critical point of the problem, nor any other spectral information, at variance with alternative schedule optimization 
approaches~\cite{Roland_PRA2002, Polkovnikov_PRL2008} which are explicitly tailored from the known critical bottleneck of the QA evolution. 

As a possible generalization, we mention that interesting results, which will be the subject of a separate publication~\cite{glen_unpub2}, are obtained 
when the QAOA technique, in its VQCS variant~\cite{Ho_SciPost2019}, is applied to preparing the quantum ground state of 
$\Ham_{T}=\Ham_z+g\Ham_x$, again in the quantum Ising chain case. 
Here again, perhaps surprisingly, the critical point $g=1$ appears to play a prominent role~\cite{glen_unpub2}.
Applications to the infinitely connected $p$-spin Ising ferromagnet~\cite{Wauters_PRA17} are also under way and will be reported elsewhere~\cite{glen_unpub3}.

Having illustrated the construction of the optimal digitized-QA protocol in the transitionally invariant Ising chain,
one should explore the generality and limitations of such construction. 
In future investigations, we will use more sophisticate tools, such as as DMRG\cite{Schollwock_AnnPhys2011}, to analyze more general systems (e.g. not integrable). 
However, the present work still allows us to identify some key properties of the problem Hamiltonian, that played a central role in deriving the results.
The locality and the $\mathbb{Z}_2$ symmetry of the Hamiltonian are key ingredients in the derivation of the variational bound on the QAOA's performance. 
The infinitely connected $p$-spin Ising ferromagnet~\cite{Wauters_PRA17} might serve as a good testing ground for these ideas, as locality is destroyed, 
the transition is second-order for $p=2$, but becomes first-order for $p\ge 3$, and and the spin inversion symmetry is lost for odd $p$. 

The properties of the translationally invariant Ising QAOA lanscape, such as the degeneracy of optimal solutions, also facilitated the classical search for an optimal digitized QA protocol. However, the QAOA landscape of disordered systems is extremely rugged, and the search for global optimal solution in complex landscapes is in itself a computationally hard problem. Machine Learning is a promising tool to cope with the complexity of QAOA landscapes~\cite{Bukov_PRX18, Troyer_PRA16, Crooks_arXiv2018}. 
The application of Machine Learning ideas may allow to extend the construction of optimal digital adiabatic protocols to disordered systems.
 


%
      
\section*{ACKNOWLEDGMENTS}        
We acknowledge fruitful discussions with L. Arceci and M. Wauters. 
Research was partly supported by EU Horizon 2020 under ERC-ULTRADISS, Grant Agreement No. 834402.
RF and GES acknowledge that their research has been conducted within the framework of the Trieste Institute for Theoretical Quantum Technologies (TQT).

\appendix
\section{Restoring translational invariance with ABC} \label{app:ABC_to_PBC}
We consider a reduced Ising chain of $\Nred=2\Ptrot+2$ spins with anti-periodic boundary conditions (ABC). 
As in the main text, we number the spins using their position relative to the reduced chain.
\begin{widetext}
In this section we give a proof of the identity
\begin{eqnarray} \label{eqn:antiperiodic_average}
\bra{\widetilde{\psi}_\Ptrot(\bgamma, \bbeta)} \PauliSigma^z_{j_s} \PauliSigma^z_{j_s+1} \ket{\widetilde{\psi}_{\Ptrot}(\bgamma, \bbeta)} &=&
\frac{1}{\Nred} \bra{\widetilde{\psi}_{\Ptrot}(\bgamma, \bbeta)} 
\bigg( \sum_{j=1}^{\Nred-1} \PauliSigma^z_{j}\PauliSigma^z_{j+1} - \PauliSigma^z_{\Nred} \PauliSigma^z_{1} \bigg) 
\ket{\widetilde{\psi}_{\Ptrot}(\bgamma, \bbeta)}  \;,
\end{eqnarray}
where $j_s$ is any internal lattice site, and the expression of $\ket{\widetilde{\psi}_{\Ptrot}(\bgamma, \bbeta)}$ (see Eq.~\eqref{eqn:psi_tilde_qaoa}) is 
\begin{equation}\label{eqn:psi_tilde_qaoa_bis}
 \ket{ \widetilde{\psi}_\Ptrot(\bgamma,\bbeta) } = \Tprod{\Ptrot}_{m=1} \ee^{-i \Hred_x \beta_m} \ee^{-i \Hred^{\smallm}_z \gamma_m} \; | \widetilde{\psi}_0\rangle \;,
\end{equation}
\end{widetext}
with $\ket{\widetilde{\psi}_0} = |+\rangle^{\otimes \Nred}$ and the reduced chain Hamiltonians given by:
\begin{eqnarray}
\Hred^{\smallm}_z + \Nred &=& 
\sum_{j=1}^{\Nred-1} \PauliSigma^z_{j}\PauliSigma^z_{j+1} - \PauliSigma^z_{\Nred} \PauliSigma^z_{1} \\
\Hred_x &=& 
-\sum_{j=1}^{\Nred} \PauliSigma^x_j\,.
\end{eqnarray}
Notice that the expression appearing on the right-hand side of Eq.~\eqref{eqn:antiperiodic_average} coincides, apart from the constant $\Nred$, with 
$\Hred^{\smallm}_z$, the reduced spin chain Hamiltonian with ABC ($J_b=-1$) introduced in Eq.~\eqref{eqn:Hred_z}. 
In the main text we used the identity in Eq.~\eqref{eqn:antiperiodic_average} to derive the expression for the residual energy given in Eq.~\eqref{eqn:eres_final}.

The key to the proof of Eq.~\eqref{eqn:antiperiodic_average} is showing that there exits a unitary 
``anti-periodic" translation transformation $\Ttwistop$ that is a symmetry of the Hamiltonians. 
Given the usual translation operator $\Top$:
\begin{eqnarray}\label{eqn:translation_sym_pbc_1}
    \Top^\dagger \bpauli_{j} \Top &=& \bpauli_{j+1} \hspace{10mm}  \mbox{ for } j\neq \Nred\\
    \Top^\dagger \bpauli_{\Nred} \Top &=& \bpauli_{1}\,,\label{eqn:translation_sym_pbc_2}
\end{eqnarray}
we define the anti-periodic translation operator $\Ttwistop$ to be the unitary transformation obtained by composing the standard translation 
$\Top$ with a flip of the first spin: $\Ttwistop \equiv \Top\PauliSigma^x_{1}$. 
The action on the spin operators induced by $\Ttwistop$ is
\begin{eqnarray} \label{eqn:translation_sym_abc_1}
\Ttwistop^\dagger \bpauli_{j} \Ttwistop &=& \PauliSigma^x_{1}\bpauli_{j+1}\PauliSigma^x_{1} = 
\bpauli_{j+1} \hspace{5mm} \mbox{ for } j\neq \Nred \hspace{8mm}\\
\Ttwistop^\dagger \bpauli_{\Nred} \Ttwistop &=& \pauli^x_{1}\bpauli_{1}\pauli^x_{1}
=(\PauliSigma^x_1, -\PauliSigma^y_1, -\PauliSigma^z_1)^T \;.  \label{eqn:translation_sym_abc_2}
\end{eqnarray}
Using Eq.~\eqref{eqn:translation_sym_abc_1} and Eq.~\eqref{eqn:translation_sym_abc_2}, a 
straightforward computation shows that 
\begin{equation}
\Ttwistop^\dagger \Hred^{\smallm}_z \Ttwistop = \Hred^{\smallm}_z \;,
\end{equation}
while the invariance of $\Hred_x$ and of the inital state $\ket{\widetilde{\psi}_0}$ is trivial. 
This in turns implies the identity
\begin{eqnarray}\label{eqn:psi_tilde_traslation_invar}
\Ttwistop \ket{\widetilde{\psi}_{\Ptrot}(\bgamma, \bbeta)} &=& \ket{\widetilde{\psi}_{\Ptrot}(\bgamma, \bbeta)} \;.
\end{eqnarray}
Additionally, Eq.~\eqref{eqn:translation_sym_abc_1} and Eq.~\eqref{eqn:translation_sym_abc_2} also 
imply that $\Hred^{\smallm}_z +\Nred$ decomposes into a sum of terms obtained by applying powers of $\Ttwistop$ to   
$\PauliSigma^z_{j_s}\PauliSigma^z_{j_s+1}$:
\begin{eqnarray}\label{eqn:Hred_traslation_sum}
\Hred^{\smallm}_z +\Nred &=& \sum_{n=0}^{\Nred-1}\Ttwistop^{\dagger n} \, \PauliSigma^z_{j_s}\PauliSigma^z_{j_s+1}\,\Ttwistop^n \;.
\end{eqnarray}
\begin{widetext}
The desired equality is a direct consequence of Eq.~\eqref{eqn:psi_tilde_traslation_invar} and Eq.~\eqref{eqn:Hred_traslation_sum}. 
Indeed, one has that
\begin{eqnarray}
\bra{\widetilde{\psi}_{\Ptrot}(\bgamma, \bbeta)}\Hred^{\smallm}_z 
+\Nred\ket{\widetilde{\psi}_{\Ptrot}(\bgamma,\bbeta)} &=&
\bra{\widetilde{\psi}_{\Ptrot}(\bgamma, \bbeta)} \sum_{n=0}^{\Nred-1}\Ttwistop^{\dagger n}\,
\PauliSigma^z_{j_s}\PauliSigma^z_{j_s+1} \,\Ttwistop^n
\ket{\widetilde{\psi}_{\Ptrot}(\bgamma, \bbeta)}\nonumber\\
&=&\sum_{n=0}^{\Nred-1}\bra{\widetilde{\psi}_{\Ptrot}(\bgamma, \bbeta)} 
\PauliSigma^z_{j_s}\PauliSigma^z_{j_s+1}
\ket{\widetilde{\psi}_{\Ptrot}(\bgamma, \bbeta)}
=\Nred\bra{\widetilde{\psi}_{\Ptrot}(\bgamma, \bbeta)} 
\PauliSigma^z_{j_s}\PauliSigma^z_{j_s+1}
\ket{\widetilde{\psi}_{\Ptrot}(\bgamma, \bbeta)} \;, \hspace{10mm}
\end{eqnarray}
which proves Eq.~\eqref{eqn:antiperiodic_average}. 
\end{widetext}

\section{QAOA landscape $\eres(\bgamma,\bbeta)$} \label{app:QAOA_landscape}
This appendix contains additional useful material on the QAOA residual energy landscape $\eres_{\Ptrot}(\bgamma,\bbeta)$,
in particular concerning some of its symmetry properties.

The residual energy $\eres_{\Ptrot}(\bgamma,\bbeta)$ defines a landscape for the classical optimization routine 
(which for our implementation is the BFGS algorithm) involved in the QAOA. 
Since the shape of the landscape partly determines the hardness of finding the desired optimal values 
$\bgamma^\star,\bbeta^\star$, in this section we discuss some of its properties.

We start by recalling that the landscape is periodic in each variable $\gamma_m$ and $\beta_m$, 
with a period of $\pi/2$. Therefore, as in the main text, without loss of generality we assume that  
$\gamma_m, \beta_m\in[0,\frac{\pi}{2})$.
From the transformation properties of the Hamiltonian we get the following fundamental relations:
\begin{itemize}
\item As shown in Ref.~\onlinecite{Wang_PRA2018}, a direct consequence of the 
{\em duality}~\cite{book:Suzuki2012} of the Ising model, is that
\begin{equation}\label{eqn:duality_sym}
    \eres_{\Ptrot}(\bgamma, \bbeta) =  \eres_{\Ptrot}(\frac{\pi}{2}-\bbeta^\prime, 
    \frac{\pi}{2}-\bgamma^\prime)\,,
\end{equation}
where $\bbeta^\prime=(\beta_\Ptrot, \beta_{\Ptrot-1},\cdots,\beta_1)^T$ and  
$\bgamma^\prime=(\gamma_\Ptrot, \gamma_{\Ptrot-1},\cdots,\gamma_1)^T$.
This relation is specific to the model we are considering.
\item By applying a spin {\em flip on even sites} 
$\Parityop=\prod_{n=1}^{N_s/2}\PauliSigma^x_{2n}$ one can change the sign of the cost function 
Hamiltonian $\Parityop(\Ham_z+N_s)\Parityop=-(\Ham_z+N_s)$ --- essentially mapping the antiferromagnetic couplings to ferromagnetic ones ---
while maintaining $\Ham_x$ unchanged. 
After some straightforward algebra this implies that
\begin{equation}\label{eqn:ferro-antiferro_sym}
\eres_{\Ptrot}(\bgamma, \bbeta) = 1 -\eres_{\Ptrot}(\frac{\pi}{2}-\bgamma, \bbeta)\,.
\end{equation}
This relation holds for any bipartite graph.
\end{itemize}
From Eq.~\eqref{eqn:duality_sym} and Eq.~\eqref{eqn:ferro-antiferro_sym} one can then derive
\begin{eqnarray}
\eres_{\Ptrot}(\bgamma, \bbeta) &=& \eres_{\Ptrot}(\frac{\pi}{2}-\bgamma, 
\frac{\pi}{2}-\bbeta)\label{eqn:sym_origin}\\
\eres_{\Ptrot}(\bgamma, \bbeta) &=& \eres_{\Ptrot}(\bbeta^\prime, 
\bgamma^\prime)\label{eqn:sym_duality}\\
\eres_{\Ptrot}(\bgamma, \bbeta) &=& \eres_{\Ptrot}(\frac{\pi}{2}-\bbeta^\prime, 
\frac{\pi}{2}-\bgamma^\prime)\label{eqn:sym_origin_and_duality}\\
\eres_{\Ptrot}(\bgamma, \bbeta) &=& 1-\eres_{\Ptrot}(\frac{\pi}{2}-\bgamma, \bbeta)\\
\eres_{\Ptrot}(\bgamma, \bbeta) &=& 1-\eres_{\Ptrot}(\bgamma, \frac{\pi}{2}-\bbeta)\\
\eres_{\Ptrot}(\bgamma, \bbeta) &=& 1-\eres_{\Ptrot}(\bbeta^\prime, \frac{\pi}{2}-\bgamma^\prime)\\
\eres_{\Ptrot}(\bgamma, \bbeta) &=& 1-\eres_{\Ptrot}(\frac{\pi}{2}-\bbeta^\prime, \bgamma^\prime)
\end{eqnarray}
In particular Eq.~\eqref{eqn:sym_origin} (inversion symmetry) and 
Eq.~\eqref{eqn:sym_origin_and_duality} (duality symmetry) 
define two  independent symmetries of the landscape. 
The symmetry group is isomorphic to $\mathbb{Z}_2\times\mathbb{Z}_2$. 
However, although the landscape is symmetric, the optimal solution in 
general may break the symmetry. Indeed, since the only point that satisfies inversion symmetry 
is the origin, this must be broken by the minimization. Numerical results suggest that, instead, the 
duality symmetry is preserved. 
We extended the work in Ref.~\onlinecite{Wang_PRA2018} and 
verified that global minima lie in the $\bbeta'=\bgamma$ manifold up to $\Ptrot=128$.

\section{Effective Hamiltonian for digital evolution} \label{app:Heff}
\begin{figure}
    \includegraphics[width=0.8\columnwidth]{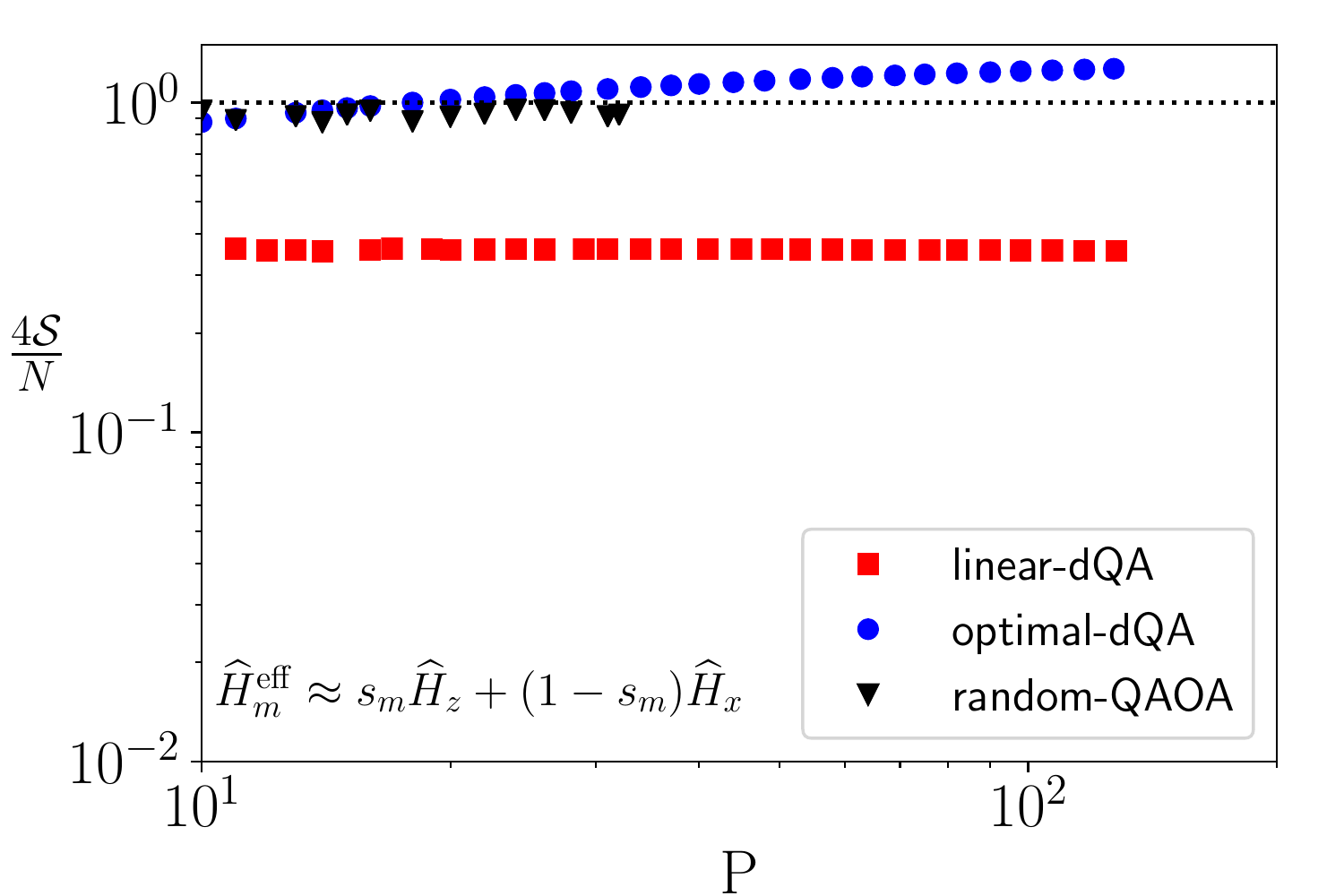}
    \caption{Approximated average Shannon entropy $\calS_{ \bgamma,\bbeta}$ defined in Eq.~(35), for various schedules. 
   The results were obtained using the second order approximation of $\Ham^{\effective}$ given in Eq.~\eqref{eqn:Heff_2nd_approx}. }
    \label{fig:QS_adiabaticity_approxHeff}
\end{figure}

In Sec.~IVA we analyzed the adiabaticity of the digital evolution operator $\Uevol_m=\Uevol(\gamma_m,\beta_m)$.
To make the analogy with the continuous-time evolution stronger, we now introduce an effective Hamiltonian $\Ham^{\effective}_m$ and a time 
discretization $\Delta t_m>0$ which satisfy
\begin{eqnarray}\label{eqn:Heff}
\ee^{-i \frac{\Delta t_m}{\hbar} \Ham^{\effective}_m} &\stackrel{\mathrm{def}}{=}& \Uevol_m =\ee^{-i \beta_{m} \Ham_{x} }\ee^{-i\gamma_{m} \Ham_z}\;,
\end{eqnarray}
where an additional condition on the spectrum of $\Ham^{\effective}_m$ must be imposed to ensure an unambiguous definition of the logarithm  
({\em e.g.} spectrum bounded in $[-\frac{\pi \hbar}{\Delta t_m},\frac{\pi \hbar}{\Delta t_m}]$). 
Clearly, the definitions given in Eq.~\eqref{eqn:Heff} is closely related to the lowest-order Trotter decomposition in Eq.~(4) of the main text. 
Indeed, under the assumption $\gamma_m, \beta_m \ll J^{-1}$ and $\gamma_m + \beta_m> 0$, we can use 
\begin{eqnarray}\label{eqn:Heff_2nd_approx}
    \Delta t_{m} &=& \hbar (\gamma_m+\beta_m) > 0 \hspace{5mm} \mbox{and} \hspace{5mm} s_{m} = \frac{\gamma_m}{\gamma_m+\beta_m} \nonumber\\
    \Ham^{\effective}_m &=&  s_m \, \Ham_z + (1-s_m) \, \Ham_x + \calO((\Delta t_m)^2)
\end{eqnarray}
to approximately describe the discrete dynamics of the system.
Unfortunately, this assumption does not hold for a generic digital evolution and an indiscriminate application of Eq.~\eqref{eqn:Heff_2nd_approx} 
may lead to incorrect results. 
In particular, the regular schedule is such that $\gamma_m^\regular, \beta_m^\regular\approx J^{-1}$ for most values of $m$, so that Eq.~\eqref{eqn:Heff_2nd_approx} 
cannot be used to get $\Ham^{\effective}_m$. One must then use other methods to compute it. 
In Fig.~\ref{fig:QS_adiabaticity_approxHeff} we show that the average Shannon entropy, $\mathcal{S}$ defined in Eq.~(35) of the main text, when 
computed using the approximate effective Hamiltonian given in Eq.~\eqref{eqn:Heff_2nd_approx} does not correctly signal the adiabaticity of the schedule 
$(\bgamma^\regular,\bbeta^\regular)$.

Although, in most cases, computing the exact expression of $\Ham^{\effective}_m$ is extremely complicated, the Jordan-Wigner pseudo-spin description 
allows us to derive an exact expression for $\Ham^{\effective}_m$ in the ordered Ising chain case. 
\begin{widetext}
In the pseudo-spin picture each $k$-vector Hilbert space evolves independently with an effective Hamiltonian given by 
\begin{eqnarray}
\Ham_{m}^{(k)} 
&=& \frac{i \hbar}{\Delta t_m}\log \left[ \ee^{-i\beta_m \Hred^{(k)}_{x} } \ee^{-i\gamma_m \Hred^{(k)}_{z}}\right] \nonumber\\
&=& \frac{i \hbar}{\Delta t_m}\log \left[
(\cos 2\beta_m + i\sin 2\beta_m\,\,\versorz\cdot\bpaulitau_k)
(\cos 2\gamma_m + i\sin 2\gamma_m\,\,\versorb_{k}\cdot\bpaulitau_k)
\right] \nonumber \\
&=& \frac{i \hbar}{\Delta t_m}\log\bigg[
 \cos (\omega^{(k)}_m \Delta t_m) +i\sin(\omega^{(k)}_m\Delta t_m) \,\,\versorom^{(k)}_m\cdot\bpaulitau_k \bigg] \nonumber \\
&=& \frac{i \hbar}{\Delta t_m}\log\bigg[ \ee^{i \Delta t_m \bomega^{(k)}_m \cdot \bpaulitau_k } \bigg] 
= -\hbar\bomega^{(k)}_m \cdot \bpaulitau_k \;,
\end{eqnarray}
\end{widetext}
where we used standard properties of the Pauli matrices, $({\bf u} \cdot \bpaulitau) ({\bf v} \cdot \bpaulitau) = {\bf u}\cdot {\bf v} + i ({\bf u}\times {\bf v}) \cdot \bpaulitau$ for
any two three-dimensional vectors ${\bf u}$ and ${\bf v}$, and we introduced an effective ``magnetic field'' $\bomega_m^{(k)}$ 
\begin{eqnarray}
    \bomega^{(k)}_m \Delta t_m &=& \cos 2\beta_m \sin 2\gamma_m \,\,\versorb_{k} + \cos 2\gamma_m \sin 2\beta_m \,\,\versorz \nonumber \\
    && -\sin 2\beta_m \sin 2\gamma_m\,\,\versorz\times\versorb_k
\end{eqnarray}
with associated unit vector $\versorom^{(k)}_m=\bomega^{(k)}_m/|\bomega^{(k)}_m|$ and the frequency $\omega^{(k)}_m=|\bomega^{(k)}_m|$, which can 
also shown to be such that:
\begin{equation}
   \cos( \omega^{(k)}_m \Delta t_m) = \cos 2\beta_m\cos 2\gamma_m-\sin 2\beta_m\sin 2\gamma_m\,\,\versorb_k\cdot\versorz  \;.
\end{equation}
%
%
We observe that the ambiguity in the logarithm has been transferred to the trigonometric functions.

To address the issue of the ``criticality'', as seen from the digital dynamics perspective, we now look for points in parameter space where the 
``effective magnetic field'' {\em vanishes}. 
\begin{widetext}
Using the fact that $\versorb_k=(-\sin k,0,\cos k)^{T}$, after rather simple algebra one can show that:
\begin{equation}
\big| \bomega^{(k)}_m \Delta t_m \big|^2 = \sin^2 2(\beta_m-\gamma_m) + (1-\cos^2 k) \sin^2 (2\beta_m) \sin^2 (2\gamma_m) +
\frac{(1+\cos k)}{2} \sin (4\beta_m) \sin (4\gamma_m) \;,
\end{equation}
where one should recall that the parameters $\beta_m$ and $\gamma_m$ can always be taken to be in the interval $[0,\frac{\pi}{2}]$. 
One can show that the only $k$ for which such a quantity can possibly vanish is $k=\pi$, and the critical parameters are such that $\beta_m\approx \gamma_m$.
Close to such a point one can expand such a quantity as:
\begin{equation}
\big| \bomega^{(k)}_m \Delta t_m \big|^2 = \sin^2 2(\beta_m-\gamma_m) + (k-\pi)^2
\left( \sin^2 (2\beta_m) \sin^2 (2\gamma_m) + \frac{1}{4}  \sin (4\beta_m) \sin (4\gamma_m) \right) \;,
\end{equation}
from which we observe that for $\beta_m=\gamma_m$ the effective field behaves in the usual Ising-like fashion:
\begin{equation}
\big| \bomega^{(k)}_m \Delta t_m \big| \sim | k-\pi | |\sin 2\gamma_m| + \cdots 
\end{equation}
\end{widetext}
This shows that the ``digital criticality'' is associated to $\beta_m=\gamma_m$, which translates into $s_m=\frac{1}{2}$ in terms of the $s$-parameter,
precisely as for the continuous-time case.

\bibliography{../../BiblioQIC,../../BiblioQAOA,../../BiblioQA,../../BiblioNishimori,../../BiblioQA_bis,../../BiblioQIsing} 

\end{document}